\documentclass[10pt, conference, compsocconf]{IEEEtran}
\pdfoutput=1
\usepackage{amsfonts,amssymb,amsmath,alltt,stmaryrd}
\usepackage {mathpartir,amssymb}
\usepackage{graphicx,url}
\usepackage[latin1]{inputenc}
\usepackage{flushend}

                      {\end{alltt}}
\def\ttbraces{\let\.=\nobreak\chardef\{=`\{\chardef\}=`\}\chardef\|=`\\}
\newcommand\aspfun{ASP${}_\text{fun}$\ }
\newcommand\aspfunp{ASP${}_\text{fun}$}
\newcommand{\symb}[1]{\makebox{\it #1}} 
\newcommand\dom{{\rm dom}}
\newcommand\ran{{\rm ran}\ }
\newcommand\imp\Longrightarrow

\newcommand\inda{\sim_{\alpha}}

\newcommand\indR{{=}_{R}\,}
\newcommand{\coloncolon}{\mathrel{::}}
\newcommand\loc{\ensuremath{\to_\varsigma}}
\newcommand\dist{\ensuremath{\to_\|}}

\newcommand\diststar{\ensuremath{\to_\|^*}}
\newcommand{\setN}{{\mathord{\mathbb N}}}
\newcommand\ie{i.e.\!\,, }
\newcommand\all{\forall\,}

\newtheorem{theorem}{Theorem}
\newtheorem{definition}{Definition}[section]
\newtheorem{proposition}[definition]{Proposition}

\newtheorem{lemma}[definition]{Lemma}

\newtheorem{property}[definition]{Property}

\newcommand{\da}{\delta_{\alpha}}

\begin{document}

\title{Confinement for Active Objects}
\author{%
\IEEEauthorblockN{Florian Kamm\"uller\\}
\IEEEauthorblockA{
Middlesex University, London UK and\\
Technische Universit\"at Berlin, Germany\\
Email: f.kammueller@mdx.ac.uk}
}
\maketitle
\thispagestyle{empty}

\begin{abstract}
In this paper, we provide a formal framework for the security of distributed active objects.
Active objects communicate asynchronously implementing method calls via futures.
We base the formal framework on a security model that uses a semi-lattice to enable
multi-lateral security crucial for distributed architectures.
We further provide a security type system for the programming model ASPfun of 
functional active objects.
Type safety and a confinement property are presented. ASPfun thus realizes secure down calls.
\end{abstract}
\section{Introduction}
\label{sec:intro}
Formal models for actor systems become increasingly important  for the security
analysis of distributed applications.
For example, models of organisational structures together with actors provide a basis
for the analysis of insider threats, \cite{Probst:2008:EAS:1480242.1480308,Probst.311.1}.

Active objects define a programming model similar to actors \cite{Agha:92}
but closely related to object-orientation.  An object is an {\it active object} if 
it serves as an access point to its own methods and associated (passive) objects and their 
threads.  Consequently, every call
to those methods will be from outside. These remote calls are 
collected in a list of requests.
The unit comprising the object's methods and attributes and its current requests is 
called {\it activity}.
The activity serves as a unit of distribution since it has a data space separate from its environment 
and can process requests independently.
To enable asynchronous communication between distributed active objects, the concept of 
{\it futures} -- promises for method call values -- is used.
Active objects are practically implemented in the Java API ProActive \cite{CDD:CMST06}
developed by Inria and commercialized by its spin-off ActiveEON.
Active objects are also a tangible abstraction for distributed information systems beyond just
one specific language.
ASP \cite{CHSPOPL04} is a calculus for active objects. 
ASP has been simplified into \aspfun --
a calculus of {\it functional} active objects. \aspfun is formalized
in Isabelle/HOL \cite{hkl:11} thus providing a general automated framework for 
the exploration of properties of active objects.

In this paper, we use this framework to support security specification and analysis of
active objects.
The contributions of this paper are 
(a) the formalization of a novel security model for distributed active objects 
    that supports multi-lateral security,
(b) a type system for the static security analysis for \aspfun configurations,
(c) preservation and the simple security property of confinement for well-typed configurations,
(d) and an argument that secure down calls are possible for \aspfunp.

The novel security model \cite{kam:12} is tailored to active objects as it supports
decentralized privacy specification of data in distributed entities.
This is commonly known as multi-lateral security. To achieve it we break away from 
the classical dogma of lattices of security classes and use instead semi-lattices. 
In our model, we implement {\it confinement}.
Every object can remotely access only public ($L$) methods of other activities. 
Methods can be specified as private ($H$) in an activity forbidding direct access.
All other methods of objects are assumed to be $L$, partitioning methods 
locally into $L$ and $H$. 
The security policy further forbids local information flow from $H$ to $L$.
To access an $L$-method remotely,
the containing activity must also be visible to the calling activity in 
a configuration.
In \aspfunp, this visibility relation is implemented by activity references.
In other active object programming languages, visibility could be given
alternatively by an import relation or a registry. 

In this paper, we provide an implementation of this security model in the \aspfun framework to 
illustrate its feasibility and the applicability of the \aspfun framework.

We design a security type system for \aspfun that implements a type check for a
security specification of active objects and visibility.
We prove the preservation property for type safety of the type system guaranteeing that
types are not changed by the evaluation of an \aspfun configuration.
The specification of parts of an active object as confined, or private (or $H$), 
is possible at the discretion of the user. This specification is entered as a security 
assignment into the type system;
by showing a general theorem that confinement is entailed in well-typedness, 
we thus know that a well-typed program provides confinement of private methods.
Although the confinement property intuitively suffices for security, at this point, a 
formal security proof is still missing. Moreover, implicit flows may occur.
We thus 
provide a definition of noninterference for active objects. Based on that, we prove 
that a well-typed configuration does not leak information to active objects below 
in the hierarchy of the security model, \ie multi-lateral security holds for well-typed 
configurations.

Remote method calls in \aspfun have no side-effects. Hence, secure down calls can be made.
Confinement provides that no private information is accessed remotely and side-effect freedom
guarantees that through the call no information from the caller side is leaked.
Side effects are excluded in our formal model \aspfun because it is functional 
but this can be implemented into the run-time system of other active object languages.

\subsection*{Overview}
We first review the semi-lattice for multi-lateral security 
(Section \ref{sec:sml4mls}) 
and \aspfun (Section \ref{sec:aspfun}) introducing
a running example of private sorting (Section \ref{sec:ex}).
Next, we describe how the semi-lattice model can be applied to active
objects by instantiating it for \aspfunp (Section \ref{sec:smlasp}). We 
discuss secure down calls, a distinctive feature of \aspfun enabled by 
its functional nature and that moreover does not restrict common bi-directional
communication patterns. To show the latter point, we present
how to implement the Needham-Schroeder Public Key protocol in \aspfunp.
We describe what we mean by security, i.e., the attacker model and the information 
flows between active objects through method calls (Section \ref{sec:sec}) and illustrate
their enforcement on the running example.
Following that, we present a type system for the static analysis of a configuration of 
active objects in \aspfun (Section \ref{sec:types}). Properties of this type system are 
presented  (Section \ref{sec:props}): 
(a) preservation as a standard result of type safety and (b) confinement.
We then define noninterference and multi-lateral security formally to present a soundness
theorem, i.e., well-typed configurations are multi-lateral secure.
We finish the paper with a related work section and also give some conclusions 
(Section \ref{sec:concl}).
An Appendix contains sections A \dots E with formal details, more examples, and (full) proofs.

\section{Prerequisites}
\label{sec:prereq}
\subsection{Semi-Lattice Model for Privacy}
\label{sec:sml4mls}
\begin{figure}
\vspace{-3ex}
\begin{center}
\includegraphics[scale=.1]{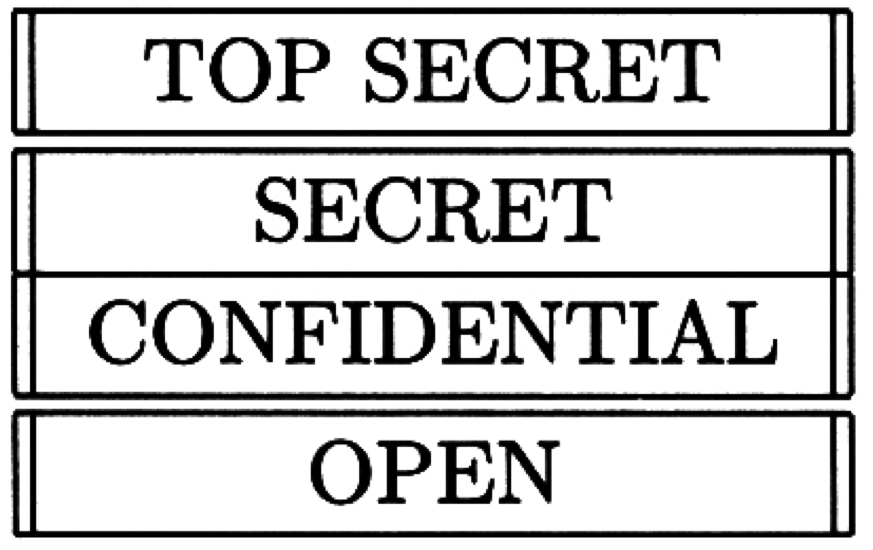}\includegraphics[scale=.1]{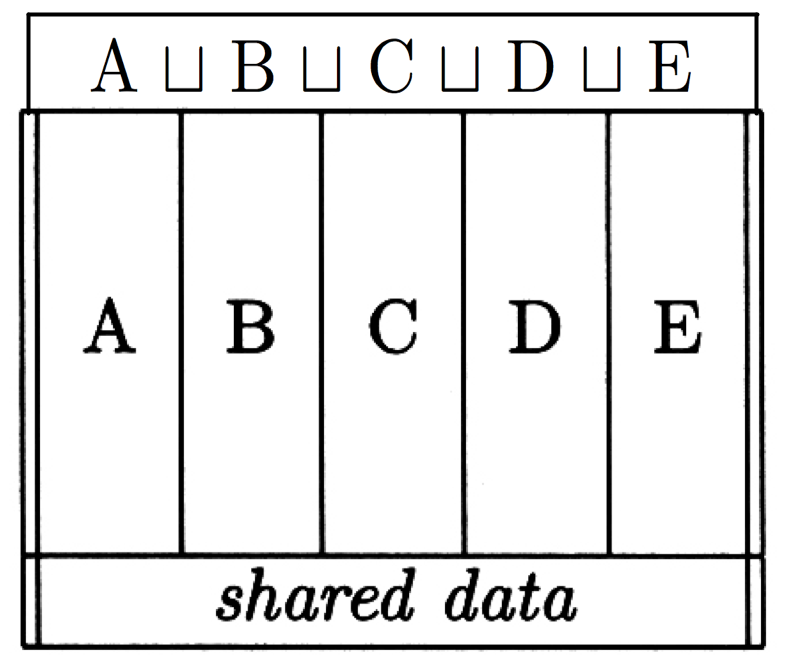}\includegraphics[scale=.1]{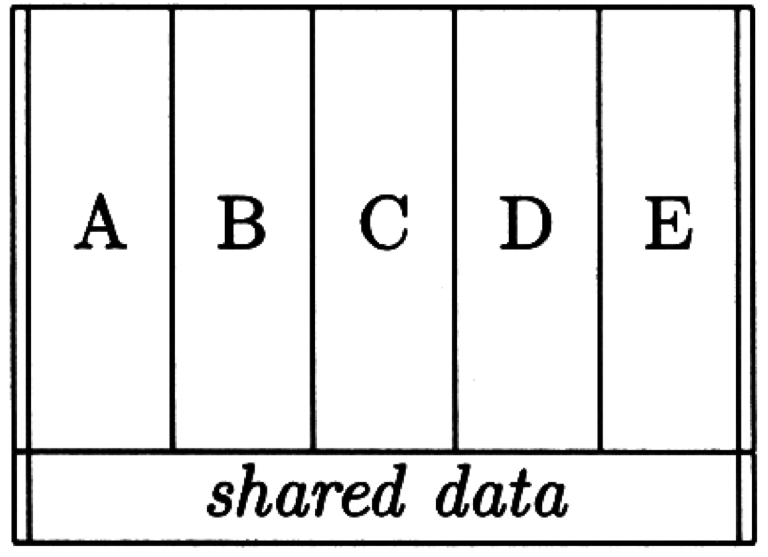}
\vspace{-2ex}
\caption{Joins enable Top control in MLS models}
\label{fig:mls}
\end{center}
\vspace{-3ex}
\end{figure}

We abstract the confinement property known from object oriented languages, e.g., 
private/public in Java,
and use it as a blueprint for a model of privacy in distributed objects. 
Consider Figure \ref{fig:mls}: multi-{\it level} security models support strict hierarchies 
like military  organization (left); multi-{\it lateral} security \cite[Ch. 8]{and:01} 
is intended to support a decentralized security world where 
parties A to E share resources without a strict hierarchy (right) thereby granting 
privacy at the discretion of each party. But lattice-based security models usually
achieve the middle schema: since a lattice 
has joins, there is a security class A $\sqcup$ B $\sqcup$ C $\sqcup$ D $\sqcup$ E 
that has unrestricted access to all 
classes A to E. For a truely decentralized multi-lateral security model 
this top element is considered harmful. 
To realize confinement, we exclude the top element by excluding joins from the lattice. 
We thereby arrive at an algebraic structure called a semi-lattice in which meets always 
exist but not joins.

\subsubsection*{Semi-Lattice}
The semi-lattice of security classes for active objects is a combination of
global and local security lattices. The two lattices are used to classify
the methods into groups and objects into hierarchies.

\subsubsection{Local Classification}
The local classification is used to control the information flow inside an
object, where methods are called and executed. 
For every active object there is the public ($L$) and a private
($H$) level partitioning the set of this active object's methods. 
The order relation of the lattice for local classification is the relation $\leq$ 
defined on $\{L,H\}$ as $\{(L,L),(L,H),(H,H)\}$.

\subsubsection{Global Classification}
The purpose of the global classification is to control the course of information flows 
between methods of globally distributed objects 
and lead their information together in a common dominating activity.
To remotely access active objects, the key is their identity 
(we use $\alpha$, $\beta$ to denote identities). 
As classes for the global lattice we use subsets of the set of all 
activity identities ${\cal I}$. These subsets of compartments
build the lattice of global 
classes, the powerset lattice $\cal{P}(I)$ over activity identities ${\cal I}$.
\[ (\cal{P}(I), \cap, \cup, \subseteq, \varnothing, I) \]
In a concrete configuration, the global class label of an activity is the set of
activity identities to which access is granted. For example, with 
respect to the Hasse diagram in Figure \ref{fig:slmmls}, an object at global level 
$\{\alpha, \beta\} \in {\cal P(I)}$ can access any
part (method) of an object labeled as $\{\beta\}$ or $\{\alpha\}$ or $\{\}$ 
but only if this part is additionally labeled as $L$.
Vice versa an object at level $\{\alpha\}$ can neither access $L$ nor $H$ parts of objects at level
$\{\beta\}$ nor any parts at level $\{\alpha, \beta\}$ but only $L$ parts at level $\{\}$. Thus the
classification of parts of an active object needs to combine labels.

\subsubsection{Combination of Lattices}
\label{sec:comblatt}
The security model of the semi-lattice needs to combine the local and global classification scheme.
As result, a {\it security class} is a pair of local and global class $(S, \delta)$.
We want to impose confinement of methods in order to realize multi-lateral security with 
our model. Thus, we have to define the combination of the two constituting lattices such that its
order relation corresponds to a multi-lateral information flow relation. I.e., private methods 
of an object are not accessible by any other than the object itself.

\begin{figure}
\vspace{-3ex}
\begin{center}
\includegraphics[scale=.2]{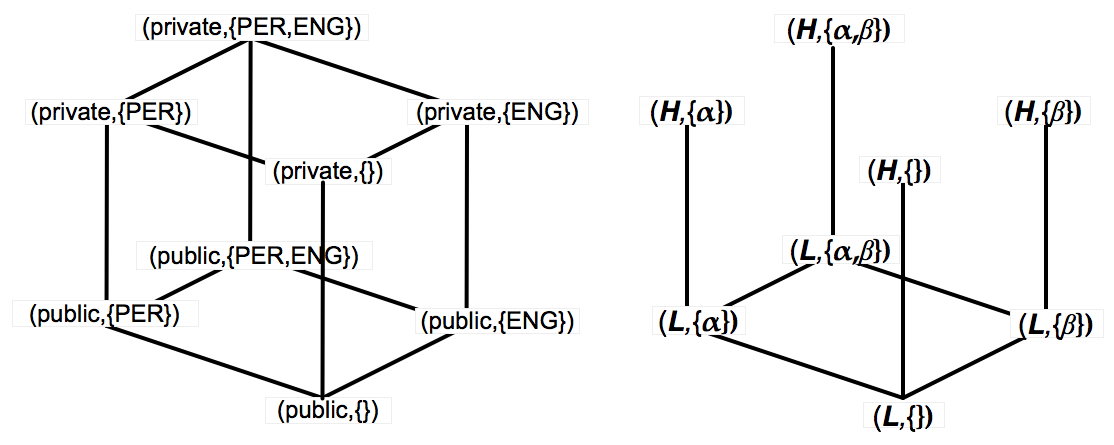}
\end{center}
\vspace{-3ex}
\caption{Taking the top off MLS lattice (left) leads to semi-lattice (right).
\label{fig:slmmls}}
\vspace{-3ex}
\end{figure}

Consequently, the new order for security classes is defined 
as follows.
The combined 
security class ordering for active objects is defined 
such that a method class $(H, \delta)$ dominates $(L, \delta)$ and also $(L, \delta')$ for all 
$\delta' \subseteq \delta$ 
but no other $(X, \delta_0)$ dominates  $(H, \delta)$. 
The combination of local and global types into pairs gives a partial order 
\[ \symb{CL} \equiv ( \{L,H\} \times \cal{P(I)}, \sqsubseteq ) \]
with 
\[ (S_0, I_0) \sqsubseteq (S_1, I_1) \equiv 
 \left(\begin{array}{c}
     S_0 <_S S_1 \vee S_0 = S_1 = L\\
     I_0 \subseteq I_1 
 \end{array} \right)
\]
where  the vertical notation $\phi \choose \xi$ abbreviates $\phi \wedge \xi$ and
$<_S = \{ (L,H) \}$ denotes the strict ordering on the local security classes.
Consequently, 
meets exist but no joins. The
partial order {\it CL} is thus just a semi-lattice as illustrated by an example 
in Figure \ref{fig:slmmls} (right).

\subsection{Functional Active Objects: \aspfun}
\label{sec:aspfun}
\aspfun uses a slightly extended form of the simplest $\varsigma$-calculus 
from the Theory of Objects \cite{AC96a} by distributing 
$\varsigma$-calculus objects into activities. 
The calculus \aspfun is functional because method update is realized on a copy of the active
object: there are no side-effects.

\subsubsection{$\varsigma$-calculus}
Objects consist of a set of labeled methods $[l_i = \varsigma(y)b]^{i\in1..n}$. Attributes 
are considered as methods not using the parameters. 
The calculus features method call $t.l(s)$ and method update $t.l := \varsigma (y) b$ on objects where
$\varsigma$ is the binder for the method parameter $y$. Every method may also contain a ``{\it this}'' element 
representing the surrounding object. Note, that the ``this'' is usually \cite{AC96a} expressed as an additional
parameter $x$ to each method's $\varsigma$ scope but we use for this exposition literally 
{\it this} to facilitate the understanding. It is, however, important to bear in mind that 
formally {\it this} is a variable representing a copy of the current object and that this 
variable is scoped as a local variable for each object.
The {$\varsigma$-calculus is Turing complete, e.g. it can simulate the $\lambda$-calculus. 
We illustrate the $\varsigma$-calculus by our example below.

\subsubsection{Syntax of \aspfunp}
\aspfun is a minimal extension of the $\varsigma$-calculus by one single 
additional primitive, the {\it Active}, for creating an activity.
In the syntax (see Table \ref{tab:syntax}) we distinguish between underlined constructs 
representing the static syntax that may be used by a programmer, while futures and active 
object references are created at runtime.
\begin{table}[!ht]
\[
\begin{array}{l@{\,}lr}
s,t & ::= \underline{ x} &\text{variable}\\
&|~ \underline{this} & \text{ generic object reference}\\
&|~ \underline{[l_j = \varsigma(y_j)t_j]^{j\in1..n}}&(\forall j,\,\symb{this}\neq y_j)\text{ object definition}\\
&|~\underline{s.l_i(t)} &(i \in1..n) \text{ method call}\\
&|~\underline{s.l_i:=\varsigma(y)t} &(i \in 1..n, \symb{this} \neq y)\text{ update}\\
&|~\underline{\symb{Active}(s)}& \text{ Active object creation}\\
&|~\alpha&\text{ active object reference}\\
&|~f_i&\text{ future}
\end{array}
\]
\vspace{-3ex}
\caption{\aspfun syntax}
\label{tab:syntax}
\vspace{-3ex}
\end{table}
We use the naming convention $s, t$ for $\varsigma$-terms, $\alpha$,
$\beta$ for active objects, $f_k, f_j$ for futures, $Q_\alpha, Q_\beta$ for request queues.

\subsubsection{Futures}
A {\it future} can intuitively be described as a promise for the result of a method call.
The concept of futures has been introduced in Multilisp \cite{Halstead85} and 
enables asynchronous processing of method calls in distributed applications:
on calling a method a future is immediately returned to the caller enabling the
continuation of the computation at the caller side. Only if the method call's value
is needed, a so-called wait-by-necessity may occur.
Futures identify the results of asynchronous method invocations to an
activity. 
Technically, we can see a future as a pair consisting of a future {\it reference} and a future {\it value}.
The future reference points to the future value which is the instance
of a method call in the request queue of a remote activity. 
In the following, we will use future and future {\it reference} synonymously for simplicity.
Futures can be transmitted between activities. Thus different activities 
can use the same future.

\subsubsection{Configuration}
A {\it configuration} is a set of activities
\[
  C::=\alpha_i[(f_{j}\mapsto s_{j})^{j\in I_i}\,,\,t_i]^{i\in1..p} 
  \]
where $\{I_i\}$ are disjoint subsets of  $\setN$. 
The unordered list $(f_{j}\mapsto s_{j})^{j\in I_i}$ represents the request queue, $t_i$
the active object, and $\alpha_i \in \dom(C)$ the activity reference.
A configuration represents the ``state''of a distributed system by the current parallel activities. 
Computation is now the state change induced by the evaluation of method calls in the request queues
of the activities.
Since \aspfun is functional, the {\it local} active object 
does not change -- it is immutable -- but the configuration is changed {\it globally}
by the stepwise computation of requests and the creation of new activities.

The constructor $\symb{Active}(t)$ activates the object $t$ by creating a new activity 
in which the object $t$ becomes active object. 
Although the active object of an activity is immutable, an update operation on
activities is provided. It performs an update on a freshly created copy
of the active object placing it into a new activity with empty request queue; 
the invoking context receives the new activity reference
in return. If we want to model operations that change active objects, we can do so using the
update. Although the changes are not literally performed on the original objects, a state change
can thus be implemented at the level of configurations (for examples see \cite{hkl:11}). 
Efficiency is not the goal of \aspfun rather 
minimality of representation with respect to the main decisive language features of 
active objects while being fully formal.


\subsubsection{Results, Programs and Initial Configuration}
A term is a result, \ie a totally evaluated term, if it is either an
object (like in \cite{AC96a}) or an activity reference. 
We consider results as values.

In a usual programming
language, a programmer does not write configurations but usual programs
invoking some distribution or concurrency primitives (in \aspfun
\emph{Active} is the only such primitive). This is reflected by the \aspfun syntax given 
above. A ``program'' is a term $s_0$ given by this static
syntax (it has no future or active object reference and no free variable). In order to be
evaluated, this program must be placed in an initial
configuration. The initial configuration has a single activity with a 
single request consisting of the user program:
\[
\begin{array}{l}
  \symb{initConf}(s_0) = \alpha[f_0\mapsto s_0,[]]
\end{array}
\]
Sets of data that can be used as {\it values} are indispensable if we want to 
reason about information flows. In \aspfunp, such values can be represented as 
results (see above) to any configuration either by explicit use of some corresponding
object terms or by appropriate extension of the initial configuration that leads to
the set-up of a data base of basic datatypes, like integers or strings. 

\subsubsection{Informal Semantics of \aspfunp}
Syntactically, \aspfun merely extends the $\varsigma$-calculus by a second parameter for methods
(the first being {\it this})
and the \emph{Active} primitive but the latter gives rise to a completely new semantic layer for
the evaluation of distributed activities in a configuration.

{\it Local} semantics (the relation $\loc$) and the 
{\it parallel} (configuration) semantics (the relation $\dist$)
are given by the set of reduction rules 
informally described as follows (see Appendix C for the formal semantics). 
\begin{itemize}
\item {\sc call, update, local}: the local reduction relation {$\loc$} is based on the $\varsigma$-calculus.
\item {\sc active}: {$Active(t)$} creates a new activity {$\alpha$}, 
  with {$t$} as its active object, global new name {$\alpha$}, and initially no futures; 
  in \aspfun notation this is {$\alpha[\varnothing, t]$}.
\item {\sc request, self-request}: a {\it method call} {$\beta.l(t)$} creates a new future
  {$f_k$} for the method $l$ of active object {$\beta$} placing the resulting future value
  onto $\beta$'s request queue; the future $f_k$ can be used to refer to the future value 
  $\beta.l(t)$ at any time.
\item {\sc reply}: {\it returns result}, \ie replaces future {$f_k$} by the
      referenced result term, \ie the future value resulting from some $\beta.l(t)$.
\item {\sc update-ao}: {\it active object update} creates a copy of the active object
      and updates the active object of the copy -- the original remains the same (functional active objects are {\it immutable}).
\end{itemize}
\subsection{Running Example: Private Sorting} 
\label{sec:ex}
As an example for a standard program consider the implementation of 
quicksort as an active object $\chi$ illustrated in Figure \ref{fig:qsort}. 
The operations we use are $::$ for list cons,
$@$ for list append, $\#$ for list length, $hd$ for the list head, and a let construct 
(see \cite{hkl:11} for details on their implementation). 

\noindent%
\begin{small}
\[
 \begin{array}{l}
   \chi \big[\varnothing, \\
   \quad [ \text{qsort} = \varsigma(y)\ \text{if}\ y = []\ \text{then}\ [] \\
   \ \qquad \qquad \qquad \ \ \text{else let}\ (a::l) = y \\
   \ \qquad \qquad \qquad \qquad \qquad (l_1,l_2) = \text{{\it this}.part}\ (a, l) \\
   \ \qquad \qquad \qquad \qquad \qquad\ l_1' = \text{if}\ \#l_1 \mathord{\leq} 1\ \text{then}\ l_1\ \text{else {\it this}.qsort}(l_1) \\
   \ \qquad \qquad \qquad \qquad \qquad\ l_2' = \text{if}\ \#l_2 \mathord{\leq} 1\ \text{then}\ l_2\ \text{else {\it this}.qsort}(l_2) \\
   \ \qquad \qquad \qquad \qquad \ \ \text{in}\ l_1' @ [a] @ l_2'\ \\
   \ \qquad \qquad \qquad \ \ \text{end},\\
   \ \quad \text{part} = \varsigma(p,y)\ \text{if}\ y = []\ \text{then}\ ([],[])\\
   \ \qquad \qquad \qquad \quad\, \text{else let}\ (a\mathord{::}l) = y \\
   \ \qquad \qquad \qquad \qquad \qquad \ \, (l_1,l_2) = \text{{\it this}.part}\ (p,l) \\
   \ \qquad \qquad \qquad \qquad \quad \ \text{in if}\ p \mathord{<} a.\text{ord then}\ (l_1, a\mathord{::}l_2)\ 
                                             \text{else}\ (a\mathord{::}l_1,l_2)\ \\
   \  \qquad \qquad \qquad \quad\, \text{end}\\
   \quad ]\big]
  \end{array}
\]
\end{small}

The quick sort algorithm in $\chi$ is parametric over a method 
``ord'', a numerical value, that is used in method ``part''. This method ord  is assumed to 
be available uniformly in the target objects  contained in the list that shall be sorted. We 
omit the parameter to calls of ord because it is unused, i.e., the empty object $[]$.

\begin{figure}
\vspace{-5ex}
\begin{center}
\includegraphics[scale=.35]{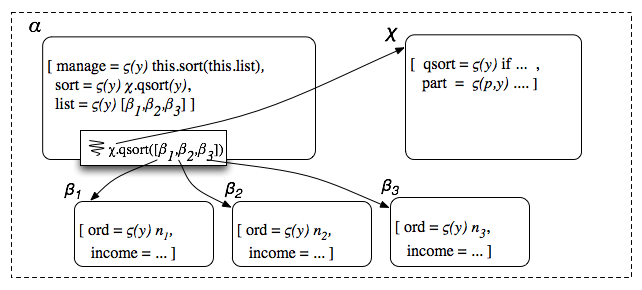}
\end{center}
\vspace{-3ex}
\caption{Three active objects $\beta_1, \beta_2, \beta_3$ in controller $\alpha$'s list.}
\label{fig:qsort}
\vspace{-5ex}
\end{figure}

The following controller object $\alpha$ holds a list of active objects (for example
$[\beta_1, \beta_2, \beta_3]$ in Figure \ref{fig:qsort} but generally arbitrary thus represented
as \dots below). 
Controller $\alpha$ uses the quick sort algorithm provided by $\chi$ to sort this list on execution
of the manage method.

\begin{small}
\[
 \begin{array}{ll}
   & \alpha \big[\varnothing, [ \text{manage} = \varsigma(y) \text{{\it this}.sort({\it this}.list)}, \\
  & \  \qquad\ \text{sort} = \varsigma(y)\ \chi.\text{qsort}(y),\\
  & \  \qquad\ \text{list} = \ldots ] \big]
 \end{array}
\]
\end{small}

The target objects contained in $\alpha's$ list (omitted) are active objects of the
kind of $\beta$ below. Here, the $n$ in the body of method ord is an integer specific to $\beta$
and the field income shall represent some private confidential data in $\beta$. 

\begin{small}
\[
 \beta \big[\varnothing,[ \text{ord} = \varsigma(y) n, \text{income} = \ldots ] \big]
\]
\end{small}

If active objects of the kind of $\beta$ represent principals in the system,
it becomes clear what is the privacy challenge: the controller object $\alpha$ should be able to
sort his list of $\beta$-principals without learning anything about their
private data, here income.

\section{Semi-Lattice Model for \aspfun}
\label{sec:smlasp}
As a proof of concept, we show that the calculus of functional active objects
\aspfun gives rise to a fairly straightforward implementation of the security semi-lattice
by mapping the concepts of the security model onto language concepts as follows.
\begin{itemize}
\item The global class ordering on sets of activity identities corresponds to the
      sets of activity references that are accessible from within an activity. We name this
      accessibility relation visibility (see Definition \ref{def:vis}). It is a consequence
      of the structure of a configuration thereby at the discretion of the configuration programmer.
\item The local classification of methods into public $L$ and private $H$ methods is specified
      as an additional security assignment mapping method names to $\{L,H\}$ at the discretion of the
      user.
\item Based on these two devices for specifying and implementing a security policy with active objects
      we devise as a practical verification tool a security type system for \aspfunp. The types of this
      type system correspond quite closely to the security classes of the semi-lattice defined in Section
      \ref{sec:sml4mls}: object types are pairs of security assignment maps and global levels.
\end{itemize}

\subsection{Assigning Security Classes to Active Objects}

\subsubsection*{Visibility}
We define visibility as the ``distributed part'' of the accessibility 
within a configuration. It derives from the activity references and
thus represents the global security specification as programmed into 
a configuration.

\begin{definition}[Visibility]
\label{def:vis}
Let $C$ be a configuration with a security specification
{\it sec} partitioning the methods of each of $C$'s active objects locally into $H$ and $L$
methods.
Then, the relation $\leq_{VI}$ is inductively defined on activity references by the following two cases.
\[
\begin{array}{lcl}
  \left(\begin{array}{c}
   \beta[Q_\beta, [l_i = \varsigma(y) t_i]^{i\in 1..n}] \in C \\
    \symb{sec}(l_i) = L \wedge t_i = E[\alpha] 
  \end{array}\right)
& \Rightarrow & \alpha \leq_{VI} \beta\\
   \left(\begin{array}{c}
    \beta[Q_\beta, [l_i = \varsigma(y) t_i]^{i\in 1..n}] \in C\\
     \symb{sec}(l_i) = L \wedge t_i = E[\gamma] \wedge \alpha \leq_{VI} \gamma 
   \end{array}\right)
& \Rightarrow & \alpha \leq_{VI} \beta
\end{array}
\]
We use the vertical notation $\phi \choose \xi$ to abbreviate $\phi \wedge \xi$; 
for context variable $E$ see Appendix C.
We then define the relation called {\em visibility} $\sqsubseteq_C^{{\symb{sec}}}$ 
as the {\it reflexive transitive closure} over $\leq_{VI}$ for any $C$, {\it sec}. \hfill$\Box$
\end{definition}
We denote the {\it visibility range} using Definition \ref{def:vis} as
$VI_{\symb{sec}}(\alpha,C) \equiv \{ \beta \in \dom(C) \mid \beta \sqsubseteq_C^{{\symb{sec}}} \alpha \}$.
The visibility relation
extends naturally to a relation $\sqsubseteq$ on global levels: every activity $\alpha \in C$
may be assigned the global level corresponding to the union of all its visible activities
$VI_{\symb{sec}}(\alpha,C)$.
This relation is a subrelation of the subset relation on the powerset of activity identities
introduced before and thus also a partial order. 
We use it as the semantics of the subtype relation in Section \ref{sec:types}.

\subsubsection*{Assigning Security Classes to Example}
To illustrate how activities are labeled in the semi-lattice model using visibility, 
consider the running example above 
where we assume the list in controller $\alpha$ to contain
various active object references $[\beta_0, \ldots, \beta_n]$.
We assign to each activity the global class containing its own identity and those of all 
its visible activities. 
For our example, the {\it global} class of controller $\alpha$ would 
be the following. 
\[\delta_{\alpha} = \{ \alpha\} \cup \delta_\chi \cup \bigcup_{i=0..n}{\rm \delta_{\beta_i}}\]
The global classes $\delta_{\beta_i}$ of the $\beta_i$ objects and $\delta_\chi$ in turn
contain all their visible objects' classes.
Thus, the global classes are ordered $\delta_{\beta_i} \subseteq \delta_\alpha$ for all $i$ and 
$\delta_\chi \subseteq \delta_\alpha$. The security classification of methods assigns
pairs of global classes and local levels to method names, for example,
${\rm ord}_{\beta_i} \mapsto (L, \delta_{\beta_i})$ and ${\rm income}_{\beta_i} \mapsto (H, \delta_{\beta_i})$.

\subsubsection*{Practical Classification of Objects}
The pairs $(S, \delta)$ in the partial order {\it CL} (see Section \ref{sec:comblatt})
are the security classes for methods of active objects. 
The semi-lattice is actually defined as a partial order on object methods rather than objects.
To classify objects we consider only the global part of the classification, i.e., the second
$\delta$ component because all methods of an active object have this $\delta$ in common.
Intuitively, this factorization corresponds to drawing objects as borders into the
semi-lattice structure (see Figure \ref{fig:slmmlsobj}). These borders represent the confinement zone
of an active object.  
\begin{figure}
\vspace{-3ex}
\begin{center}
\includegraphics[scale=.2]{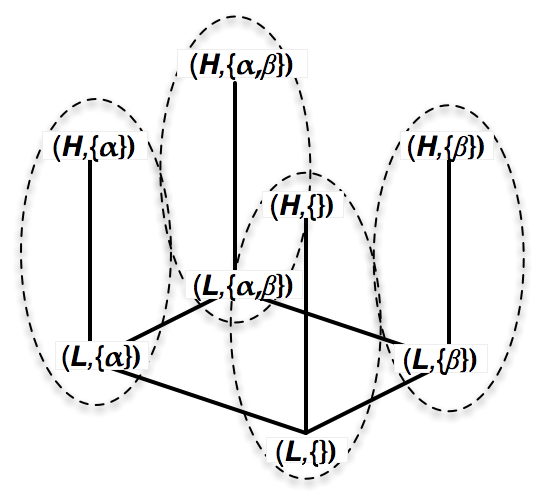}
\end{center}
\vspace{-3ex}
\caption{
Tentatively drawing in object classes as confinement zones.
\label{fig:slmmlsobj}}
\vspace{-3ex}
\end{figure}

Formally, we consider an object class to be the factorization 
$([l_i \mapsto S_i], \delta)$: a pair of a security assignment
to $\{L,H\}$ for each method $l_i$ of an object and the object's global class $\delta$ common for all parts.
An activity contains one active object but may contain various passive
objects. 
The security assignment of an active object must be defined for all contained objects
(see rule {\sc SecAss Subsumption} in Section \ref{sec:types}).



\subsection{Secure Down Calls}
In a distributed system with a nontrivial security classification of communicating objects,
secure down calls are not possible because they would violate the security
policy of ``no-down-flows'' of information.
In general, a method call represents an information flow to the remote object
in the form of the request itself and the parameters passed; 
its response flows information back in the form of a reply.
Therefore, secure method communication is trivially restricted to objects of one class --
otherwise one direction would contradict the policy ``no-down-flows''.
This catch-22 situation can 
be overcome 
if we exclude side-effects: the requests do not leave traces in the remote object. 
In \aspfun this is given implicitly by the semantics because requests created by method 
calls in the remote object are not accessible by the remote object itself.
However, the reply may flow information up. Thus, information does flow back, i.e. up. 

As an overall result of the properties presented in this paper we can infer
that secure down calls are possible. The reasoning is as follows.
We assume as given a configuration together with a security specification {\it sec}
partitioning a portion of the methods into public ($L$) and private ($H$).
If this configuration can be type checked according to our type system, it is secure,
\ie we know it  has confinement and is noninterfering as we are going to see in Section
\ref{sec:props}. Therefore, futures can be securely used in higher
security classes, \ie method results may flow up but, since no implicit flows exist, 
information is not leaked in the process.

Side-effect freedom does permit to securely call down because the call 
leaves no visible trace.
But does this not also exclude any mutual information exchange on the same level? It might 
seem so, but fortunately,
if we have two activities that are in the same class, methods 
calls between them are possible 
permitting bidirectional information flow. As an example, an
implementation of the Needham-Schroeder public key protocol is given next.

\subsection*{Needham Schroeder Public Key Protocol (NSPK) in \aspfun}
This example illustrates that inside one security class mutual information
exchange is possible between different activities. The easiest way to illustrate this 
is to use a protocol.
We use the corrected short form of the Needham Schroeder Public Key Protocol (NSPK) 
originally published by \cite{Needham78usingencryption}.
The originally published protocol missed out the $B$ inside the encrypted message to $A$ 
in step two thereby giving rise to the well-known attack of \cite{Lowe95anattack}.

The protocol is usually written as follows using public keys
$K_A, K_B$ known globally and their secret counterparts $K_{A}^{-1}, K_{B}^{-1}$ establishing 
nonces $N_A, N_B$ in the process of authentication.
\[
\begin{array}{lcl}
A \rightarrow B & : & \{N_A, A\}_{K_B}\\
B \rightarrow A & : & \{N_A,N_B,B\}_{K_A}\\
A \rightarrow B & : & \{N_B\}_{K_B}
\end{array}
\]
In \aspfunp, the protocol is implemented as a set of methods between two activities $A$ and $B$. 
We omit 
details about decoding and keys because 
it is clear that they can be implemented and we want to highlight the communication process.

\begin{small}
\[
\begin{array}{l}
A = [ \varnothing, \\
\left[ \text{own}_{id} = \dots \right. \\
\ \text{B}_{id} = \dots \\       
\ \text{step}_1 = \varsigma(y)\\
    \qquad \qquad \text{let}\ N_A = \text{new\_nonce}\\
    \qquad \qquad\quad \symb{reply} = B.\text{step}_2(\{N_A,\symb{this}.\text{own}_{id}\}_{K_B}) \\
    \qquad \qquad\quad  (N_A',N_B',B') = K_{A}^{-1}(\symb{reply})\\
     
    \qquad \qquad \text{in}\ \text{if}\ B' = \symb{this}.B_{id} \wedge N_A'=N_A\ \\
    \qquad \qquad\quad \text{then}\ (\symb{this}.\text{knows} := N_B).\text{NA} := N_A\\
    \qquad \qquad\quad \text{else}\ \symb{this}.\text{knows} :=  \text{error},\\
  \ \text{step}_3 = \varsigma(y)\\
    \qquad\qquad \text{let}\ (N_A', N_B, B_{id}') = \{y\}_{K_A^{-1}}\\
    \qquad\qquad\quad  \text{if}\ N_A' = \symb{this}.\text{NA} \wedge B_{id}' = \symb{this}.B_{id}\ \\
    \qquad\qquad\quad \text{then}\ \{N_B\}_{K_B}\\
    \qquad\qquad\quad \text{else}\ \symb{this}.\text{knows} := \text{error},\\
  \ \text{knows} = \ldots, \\
  \ \text{NA} = \dots ]]
\end{array}
\]
\end{small}
The protocol can be executed by invoking method $A.$step$_1$ which in turn invokes
the step $B.$step$_2$ and $A.$step$_3$.
\begin{small}
\[
\begin{array}{l}
B = [\varnothing, \\
\left[ \text{own}_{id} = \dots \right. \\
\ \text{A}_{id} = \dots \\    
\ \text{step}_2 = \varsigma(y) \\
    \qquad \qquad \text{let}\ N_B = \text{new\_nonce}\\
    \qquad \qquad\quad (N_A', A_{id}') = \{y\}_{K_B^{-1}}\\
    \qquad \qquad \text{in}\ \text{if}\ \symb{this}.A_{id} = A_{id}'\\ 
    \qquad \qquad\quad \text{then}\ \text{let}\ \symb{reply} = A.\text{step}_3(\{N_A',N_B,\symb{this}.\text{own}_{id}\}_{K_A})\\
    \qquad \qquad\qquad\qquad\ N_B' = \{\symb{reply}\}_{K_B^{-1}}\\
    \qquad \qquad\qquad\quad \text{in}\ \text{if}\ N_B' = N_B \\
    \qquad \qquad\qquad\qquad \text{then}\ (\text{this}.\text{knows} := N_A).\text{NB} := N_B\\
    \qquad \qquad\qquad\qquad \text{else}\ \symb{this}.\text{knows} :=  \text{error},\\
    \qquad \qquad\quad \text{else}\ \symb{this}.\text{knows} :=  \text{error},\\
  \ \text{knows} = \ldots,  \\
  \ \text{NB} = \dots ]]
\end{array}
\]
\end{small}
 In each of the steps the nonces are created,
encrypted and tested between the method calls. If the communicated messages adhere to
the protocol, \ie the nonces and ids correspond to what has been sent in earlier steps,
the own nonces are updated into the methods $A.$NA and $B.$NB and the other's
nonces in the respective method ``knows''. Otherwise, the 
protocol failure is recorded as ``error'' in method knows. This protocol implementation 
illustrates that mutual information flows are possible locally within one security class. 
The type system that we present in the following Section \ref{sec:types}
accepts this configuration since the calls are of the same global level $\delta$.

\subsection{Security Analysis}
\label{sec:sec}
In language based security, we may use the means provided by a language to enforce security.
That is, we make use of certain security guarantees that correspond to implicit assumptions 
concerning the execution of programs. The language introduces a security perimeter because we
assume that the language compilation and run-time system are respected (below the perimeter) while
the language is responsible for the security above the perimeter by virtue of its semantics
and other language tools, e.g. static analysis by type checking.
We now describe the security goal of confidentiality addressed in this paper and elaborate
on the attacker model for active objects.

\subsubsection*{Security Goal Confidentiality}
A computation of active objects is an evaluation of a distributed set of mutually referencing
activities. Principals that use the system can observe the system only by using the system's
devices. We make the simplifying assumption that principals can be identified as activities.
Principals, objects, programs and values are thus all contained in this configuration.
There are no external inputs to this system -- it is a closed system of communicating actors.
We concentrate in this paper on confidentiality, i.e. activities should not learn anything
about private parts of other activities neither directly nor indirectly.
Integrity is the dual to this notion and we believe that it can simply be derived from our present
work by inverting the order relation.

\subsubsection*{Attacker Model}
As a further consequence to the language based approach to security, we restrict the attacker to only have
the means of the language to make his observations.
Consequently, we also consider the attacker -- as any other principal -- as being represented 
by an activity. The attacker's knowledge is determined by all active objects he sees, 
more precisely their public parts. 
If any of the internal computations in 
inaccessible parts of other objects leak information,
the attacker can learn about them by noticing differences in different runs of the same
configuration. 
Inaccessible parts of other objects must be their private methods
or other objects 
that are 
referenced in these private 
parts.
The language semantics and the additional static analysis must guarantee that under the
assumption of the security perimeter an attacker cannot learn anything about private parts.

\subsection{Information Flow Control}
\label{sec:ifcao}
Information flow control \cite{dd:77} technically uses an {\it information flow policy} 
which is given by the specification of a set of {\it security classes} to classify 
information and a {\it flow relation} on these classes that defines 
allowed information flows. System entities 
that contain information, for example variables $x$, $y$,
are bound to security classes. Any operation that uses the value of 
$x$ to calculate that of 
$y$, creates a flow of information from $x$ to $y$.
This operation is only admissible if the class of $y$ dominates the class of $x$ in the flow relation,
formally written $\delta_x \sqsubseteq \delta_y$ where $\delta_e$ denotes the class of entity $e$.
The concept of information flow classically stipulates that the security classes together with 
the flow relation as an order relation on the classes are a lattice \cite{de:76,dp:02}. We differ here
since we only require a semi-lattice.

\subsubsection*{Information Flow Control for Active Objects}
Information is contained in data values which are here either objects or activity references
(see Section \ref{sec:aspfun}).
To apply the concept of information flow control to configurations of active objects, we need to
interpret the above notions of security classes, their flow relation, and the entities that
are assigned to the security classes: we identify the classes of our security model 
as the security classes of methods and the flow relation as the semi-lattice ordering 
on these classes (see Section \ref{sec:sml4mls}). 
Flows of information local to objects are generated by local method calls between
neighboring methods of the same object. These are regulated by the local $L/H$-classification of an 
object's methods ($H$ may call $L$ {\it and} $H$ -- but $L$ only $L$). 
Global flows result from remote method calls between objects' methods.
The combined admissible flows have to be in accordance with a concrete 
configuration and its $L/H$ specification. 

\subsection{Enforcing Legal Information Flows}
To illustrate the task of controlling information flows, 
we first extend the intuition about information flow to configurations of active objects. 
An active object sees only other active objects that are directly referenced in its methods
or those active objects that are indirectly visible via public methods of visible objects.
From the viewpoint of one active object, 
information may flow into the object and out of the object.
For each direction, there are two ways how information may flow: implicit or explicit (direct) flows.
Information flows {\it explicitly into} an object by parameters passed to remote calls directed to 
the object's methods; it may also flow {\it implicitly into} the object simply if the choice
of which method is called depends on the control flow of the calling object. Similarly,
information flows {\it explicitly out} of an active object by parameters passed to remote method
calls and {\it implicitly out} of it, if the choice depends on the object's own control flow.
Some of these flows are illustrated on our running example next.

\subsubsection*{Running Example: Implicit Information Flow}
\label{sec:ifex}
We will now finally illustrate the security model on the running example showing
implicit information flows of active objects introduced above in Section \ref{sec:ex}. 
Let us assume that the implementation of the $\beta$-objects featuring in the controller's list
had the following implementation.

\begin{small}
\[
 \begin{array}{l}
    \beta \big[\varnothing, \\
    \quad [ \text{ord} = \varsigma(y)\ \text{if}\ \text{{\it this}.income}\mathord{/}10^3 \mathord{\ge} 1\  \text{then}\
            1\ \text{else}\ 0 ],\\
    \  \quad \text{income} = \ldots\\
    \quad \big]
 \end{array}
\]
\end{small}

Let us further assume that ord is again a public method and income again the private field of
$\beta$.
We have here a case of an implicit information flow. Since the guard of the if-command in ord 
depends on the private field income, effectively the order number of a $\beta$-object is 1 if the 
income of $\beta$ is more than 1000 else 0. In our security model this control flow represents
an illicit flow of information from a high level value in $\beta$ to its public parts and is thus 
visible to the remote controller. This should not be the case since $H \sqsupset$ L. 
It should thus be detectable by an information flow control analysis. We will show next how to
 detect it statically by a security type system.

\section{Security Type System}
\label{sec:types}
Before formalizing security of active objects and defining a type system that implements
rules for a static analysis, we summarize the security considerations so far and motivate the 
upcoming type system and proofs.
\subsection{Intermediate Summary, Motivation, and Outlook}
\label{sec:typesmotiv}
In a configuration of active objects we may have direct (explicit) and implicit information flows
through method calls which are controlled differently.
\begin{itemize}
\item To guarantee only legal information flows on direct calls we rely on the labeling of
   methods by $L$ and $H$ and on the global hierarchy. This corresponds to the simple security
   property of {\it confinement}: remote method calls can refer only to low methods of visible objects. 
   Confinement can be locally checked. It is decidable since it corresponds to merely looking up 
   method labels in a security assignment. 
\item We will use a program counter $PC$ that records the current security level of a method evaluation. 
    Locally, within the confinement zone of an activity, 
    accessing $H$-methods in $L$-methods may create implicit flows -- as seen in the example.
    To detect such flows and protect the confidential information from flowing out of the 
    confinement zone of the activity, the program counter records these dependencies by 
    increasing to $H$. In combination with the method labels, the $PC$ thereby allows 
    associating the calling context with the called method. Implemented into type rules,
    this enables static checking and thus controlling of information flows in evaluations 
    of configurations. 
\end{itemize}
As a security enforcement mechanism of our multilateral security model for active objects, 
we propose a security type system, \ie a rule set for static analysis of a configuration with respect
to its methods' security assignment. 
The idea of a security type system is as follows.
Not all possible programs in \aspfun are secure. In general, for example, any method can be accessed
in an active object. The purpose of a type system for security is to supply a set of simple rules defining
types of configurations enabling a static check (before run-time) whether those contain only allowed 
information flows.

The above described cases of information flows need to be implemented in the type rules such that the 
rules allow to infer a type just for secure configurations and otherwise reject them.
The first direct case of information flow is intuitively simple, as it boils down to locally looking-up 
the security level of a method before deciding whether a remote call from up in the hierarchy can be 
granted.
The ``up in the hierarchy'' is encoded in a subtype relation $\sqsubseteq$ encoding the global hierarchy
described by the visibility relation. After the presentation of the type system in this section, we 
prove in the following Section \ref{sec:props} that confinement is a security property implied by it.

How to avoid and detect implicit flows, is more subtle: the combination of a program counter $PC$
with the called method's security label grants us to combine the provenance of one call with the
security level of the call context. However, this combination needs to adhere to the security specification 
for all runs of a program and thus all possible calls in a context. The appropriate notion of security
for this is noninterference: in all runs the observable (low) parts of configurations need to look
``the same''.  Therefore,  we first introduce a notion of noninterference for active objects 
based on which we will then 
be able to express the absence of implicit flows and prove multilateral security.
The definition of noninterference and proofs of properties are contained in Section 
\ref{sec:props}. We first introduce the type system.

\subsection{Type System}
\label{sec:typesformal}
\subsubsection*{Type Formation}
We need to provide types for objects and for configurations of active objects;
the latter by mapping names of futures and activities to object types.
The two-dimensional classification of local and global security described above
translates directly into the object types of the security type system.
A type is a pair  $([l_i \mapsto S_i]^{i=1..n}, \delta_\alpha)$ where $S_i \in \{L, H\}$.
The first part $[l_i \mapsto  S_i]^{i=1..n}$ provides the partition of methods into 
public ($L$) and private ($H$) methods for the object. 
The other element $\delta_\alpha$ of an object type represents the global
classification of an object. This global level corresponds to the
classification of the object's surrounding activity $\alpha$ derived from its visibility.
We adopt the following naming conventions for variables.
$\delta$ stands for the global part of a type.
We use $A$ to denote security assignments, e.g. $A = [l_i \mapsto S_i]^{i=1..n}$.
$S_i$, or simply $S$, stands for levels $L$ or $H$.
In general, we use indexed variables to designate result values of a function, 
e.g., $S_i$ for the level value of method $l_i$ -- 
also expressed as $A(l_i)$.
We use $\Sigma$ for object types $\Sigma = ([l_i \mapsto S_i]^{i=1..n}, \delta)$.
To map an object type $\Sigma$ to its security assignment or its global part,
respectively, we use the projections ${\symb ass}(\Sigma)$ and ${\symb glob}(\Sigma)$.
We formally use a parameter $\symb{sec}$ as the parameter for the overall 
methods' security assignment of an entire configuration $C$. 
A triplet of maps is a configuration type 
$\langle \Gamma_{\symb{act}}, \Gamma_{\symb{fut}}, \symb{sec} \rangle$ 
assigning types to all activities and futures of a configuration in addition containing 
the security assignment $\symb{sec}$.

\subsubsection*{Typing Relations}
A typing judgement $T; S \vdash x: ([l_i \mapsto S_i]^{i=1..n}, \delta)$ reads: 
given type assumptions in $T$, term $x$ has type $([l_i \mapsto S_i]^{i=1..n}, \delta)$
in the context of a program counter $S \in \{L, H\}$. 
A program counter ($PC$) is a common technique in information flow control 
originating in Fenton's Data Mark Machine \cite{fen:73}. The $PC$ 
encodes the highest security level that has been reached in all possible control flows leading 
to the current control state. In a functional language, like \aspfunp, this 
highest security level of all execution paths simply is the level of the evaluation context 
for the term $x$.
Thus, the $PC$ is some $S \in \{L,H\}$ denoting the security label of the local context. 
The type environment $T$ contains types $\Sigma$ for the parameter variables
$y$ and types for the parameter {\it this} both paired with the local security 
level $S$ representing their local $PC$.

\subsubsection*{Subsumption Rules}
Subsumption means that an element of a type also has the type of its super-type. 
It is responsible for making the partial order relation on global levels a subtype relation. 
Intuitively, {\sc Glob Subsumption} says that if a term can be typed in a low context
it may as well be ``lifted'', \ie considered as of higher global level thereby enforcing 
(together with {\sc Type Call} below) that only $L$-methods can be accessed remotely.
This corresponds to the confinement property as formally shown in Section \ref{sec:conf}.
The local security class ordering is $L \leq H$ and features implicitly in the type 
system in the form of a second -- the local -- subsumption rule.
Finally, the rule {\sc SecAss Subsumption} allows the security assignment type of an object
to be extended. This rule is necessary to consider an object also as a local object
inside another (active) object adopting its security assignment.
\begin{table}[!h]
\vspace{-3ex}
\begin{mathpar}
 \inferrule [Loc Subsumption]
  {T; L \vdash x : (A, \delta)}{T; H \vdash x : (A, \delta)}   

 \inferrule [SecAss Subsumption]
  {T; S \vdash x : (A, \delta) \\ A \subseteq A'}{T; S \vdash x : (A', \delta)}

 \inferrule [Glob Subsumption]
  {T; L \vdash x : (A, \delta) \\ \delta \sqsubseteq \delta'}{T; L \vdash x : (A, \delta')}
  \end{mathpar}\centering
\vspace{-3ex}
  \caption{Subsumption rules, $A = [l_i \mapsto S_i]^{i\in1..n}$, $S \in \{L, H\}$}
  \label{tab:sectypeord}
\vspace{-3ex}
\end{table}

\subsubsection*{Object Typing} 
\begin{table}[!h]
\vspace{-3ex}
\begin{mathpar}
  \inferrule [Val Self]  
      {}{\symb{this}\!:\! \Sigma \coloncolon T; \sqcup_{i \in 1..n} S_i \vdash \symb{this}\!:\! \Sigma}

  \inferrule [Val Local]
      {}{x: \Sigma :: T; S \vdash x: \Sigma}

  \inferrule [Type Object] 
    { \forall i \in 1..n.\ \\\\ \symb{this}\!:\! \Sigma \coloncolon y\!:\! \Sigma \coloncolon T; S_i \vdash t_i : \Sigma}
    {T; \sqcup_{i \in 1..n} S_i \vdash [l_i  = \varsigma(y) t_i]^{i \in 1..n}: \Sigma }

 \inferrule [Type Call] 
    {  T; S \vdash o : \Sigma \\\\ 
       j \in 1..n \\ T; S_j  \vdash t: \Sigma}
    {T; S_j \vdash o.l_j(t) : \Sigma}

 \inferrule [Type Update]
    {T; S \vdash o: \Sigma \\\\
    j \in 1..n\\ \symb{this}\!:\! \Sigma\coloncolon y\!:\! \Sigma\coloncolon T; S_j \vdash t : \Sigma}
    {T; S \vdash o.l_j := \varsigma(y) t: \Sigma }
  \end{mathpar}\centering
\vspace{-2ex}
  \caption{Type rules for objects; $\Sigma = ([l_i \mapsto S_i]^{i \in 1..n}, \delta)$}
  \label{tab:sectype}
\vspace{-3ex}
\end{table}
The object typing rules in Table \ref{tab:sectype} 
describe how object types are derived for all possible terms of \aspfunp.
The {\sc Val}-rules state that type assumptions stacked on the type environment $T$ 
left of the turnstile $\vdash$ can be used in type judgments.
These rules apply to the two kinds of environment entries for {\it this} and for the $y$-parameter.
Since the {\it this} represents the entire object value itself, its $PC$ is
derived as the supremum of all security levels assigned to methods in it. We 
express this supremum as the join over all levels $\sqcup_{i \in 1..n} S_i$.
The other rules are explained as follows.
{\sc Type Object}: if every method $l_i$ of an object is typeable with some local type 
$S_i \in \{L, H\}$ assigned to it by the assignment component $A$ of $\Sigma$, then
the object comprising these methods is typeable with their maximal local type.
Thus, objects that contain $H$ methods cannot themselves be contained in other $L$-methods.
Otherwise, local objects containing confidential parts could be typed with {\sc Glob Subsumption}
at higher levels (see the Appendix for a ``borderline example'' illustrating this point). 
Only objects that are purely made from $L$-methods can be accessed
remotely in their entirety. Albeit this strong restriction, the {\sc Call} rule permits
selectively accessing $L$ methods of such objects (see below).
The {\it PC} guarantees that all method 
bodies $t_i$ are typeable on their given privacy level $S_i$.
The rule {\sc Type Call} is the central rule 
enforcing that only $L$ methods can be called in any object -- locally or remotely.
Initially, a call $o.l_j(t)$ can only be typed as 
$\Sigma= ([l_i \mapsto S_i]^{1=1..n}, \delta)$ for the $\delta$ of the surrounding object $o$.
Although the $PC$ in the typing of $o$ is (by {\sc Type Object}) the maximal level of all methods,
we may still call $L$-methods on objects that are typed with $PC$ as $H$. 
The $PC$ in the typing of the resulting call $o.l_j(t)$ is coerced to $S_j$, \ie the
security level assigned to the called method. This prevents $H$ methods from being 
callable remotely while admitting to call methods on objects that are themselves typed 
in a $H$-{\it PC}.
Because of the rule {\sc Glob Subsumption} any method call $o.l_j(t): (A, \delta)$ can 
also be interpreted as $o.l_j(t): (A, \delta')$ for $\delta \sqsubseteq \delta'$ 
but this is restricted to $L$ contexts: a method call typeable in an $H$ context 
cannot be ``lifted'', 
\ie it cannot be interpreted as well-typed  with $\delta'$; to prevent this,
the {\it PC} in {\sc Glob Subsumption} is $L$ thus excluding {\sc Call} instantiations
for methods $l_j$ with $S_j = H$.
{\sc Update}: an update of an object's method is possible but conservatively, \ie the types must remain the same.

\subsubsection*{Configuration Typing}
The rules for configurations (see Table \ref{tab:typeconf}) use the union of all futures of a configuration.
\begin{definition}[Future Domain]
Let $C$ be a configuration. We define the domain of all futures of $C$.
\[\qquad dom_{\symb{fut}}(C) \equiv \bigcup \{ \dom(Q) \mid  \exists\ \alpha, a.\ \alpha[Q,a]\in C\} \qquad \Box\] 
\end{definition}

The rules for configurations 
anticipate two semantic properties of futures in well formed \aspfun 
configurations. We use well-formedness of \aspfun configurations as defined in \cite{hkl:11}; 
in brief: there are no dangling references. 

\begin{property}[Unique Future Home Activity]
\label{prop:futact}
Every future is defined in the request queue of one unique activity.
\[ \forall f_k \in \dom_{\symb{fut}}(C).\ \exists! \alpha[Q,a]\in C.\ f_k \in \dom(Q) \]
We denote this unique activity $\alpha$ as $\symb{futact}_C(f_k)$. \hfill$\Box$
\end{property}

Next, every future $f_k$ in a well formed configuration $C$ is created by a call to a unique label in its home activity. 
\begin{property}[Unique Future Label]
\label{prop:futlab}
Let $\alpha[Q,a]\in C$ be the unique $\symb{futlab}_C(f_k)$. Then,
\[ \forall f_k \in \dom_{\symb{fut}}(C).\ \exists! l \in \dom(a).\ \exists t.\ a.l(t) \diststar Q(f_k).\]
We denote this unique method label as $\symb{futlab}_C(f_k)$. \hfill$\Box$
\end{property}
We omit the configuration $C$ for the previous two operators if it is clear from context.

The configuration type rules
link up types for activities and futures with the local types of terms
in active objects and request lists (see Table \ref{tab:typeconf}).

\begin{table*}[ht]
\vspace{-3ex}
\centering
\begin{mathpar}
  \inferrule [Type Active]
   {\langle \Gamma_{\symb{act}}, \Gamma_{\symb{fut}}, \symb{sec} \rangle, T; S \vdash a: \Sigma}
   {\langle \Gamma_{\symb{act}}, \Gamma_{\symb{fut}}, \symb{sec} \rangle, T; S \vdash \symb{Active}(a): \Sigma}

  \inferrule [Type Active Object Reference]
   {\beta\in\dom(\Gamma_{\symb{act}})}
   {\langle \Gamma_{\symb{act}}, \Gamma_{\symb{fut}}, \symb{sec} \rangle, T, M_\beta \vdash \beta: \Gamma_{\symb{act}} (\beta) }

  \inferrule [Type Future Reference]
   {f_k \in \dom(\Gamma_{\symb{fut}})}
   {\langle \Gamma_{\symb{act}}, \Gamma_{\symb{fut}}, \symb{sec} \rangle, T; \symb{ass}(\Gamma_{\symb{fut}}(f_k))(\symb{futlab}(f_k)) \vdash f_k: \Gamma_{\symb{fut}} (f_k)}

  \inferrule [Type Configuration]
   {\dom (\Gamma_{\symb{act}}) = \dom (C) \\ \dom(\Gamma_{\symb{fut}}) = \dom_{\symb{fut}}(C) \\
   \bigcup_{\alpha \in \dom(C)} \symb{ass}(\Gamma_{\symb{act}}(\alpha)) \subseteq \symb{sec} \\\\
      {\forall\,\alpha[Q,a]\in C.~ 
\left\{ \begin{array}[c]{@{}l@{}}
          \langle \Gamma_{\symb{act}}, \Gamma_{\symb{fut}}, \symb{sec}\rangle, 
             \varnothing; M_\alpha \vdash  a: \Gamma_{\symb{act}}(\alpha)\land\\
      \forall\, f_k \!\in\! \dom(Q).\,
               \left\{ {\begin{array}[c]{@{}l@{}}
                 \Gamma_{\symb{act}}(\alpha) = \Gamma_{\symb{fut}}(f_k)\land\\
                 \langle \Gamma_{\symb{act}}, \Gamma_{\symb{fut}}, \symb{sec}\rangle, 
                          \varnothing; \symb{ass}(\Gamma_{\symb{act}}(\alpha))(\symb{futlab}(f_k))
                                                    \vdash  Q(f_k): \Gamma_{\symb{fut}}(f_k)\\
                \end{array}}  \right.
        \end{array} \right. }}
   {\vdash C: \langle\Gamma_{\symb{act}}, \Gamma_{\symb{fut}}, \symb{sec} \rangle }

\end{mathpar}
\vspace{-2ex}
 \caption{Typing configurations; $M_\alpha = \bigsqcup_{i \in 1..n_{\alpha}} {\text{\it ass}}(\Gamma_{\text{\it act}}(\alpha))(i)$}
  \label{tab:typeconf}
\vspace{-3ex}
\end{table*}

{\sc Type Active} allows to transfer the type of an object term to its activation which
coerces the types of activities and activity references to coincide with the types
of their defining objects. This is achieved together with {\sc Type Active Object Reference} and the clause 
$\langle \Gamma_{\symb{act}}, \Gamma_{\symb{fut}}, \symb{sec}\rangle, \varnothing \vdash a: \Gamma_{\symb{act}}(\alpha)$
of {\sc Type Configuration}.

{\sc Type Future Reference} similarly assigns the types for the future references in $\Gamma_{\symb{fut}}$.
For a given activity $\alpha$,
this rule further coerces the $PC$ for the typing of $f_k$ to coincide with $A_{\alpha}(\symb{futlab}(f_k))$, 
i.e., $\alpha$'s security assignment applied to the label that leads to the instance of $f_k$.

The rule {\sc Type Configuration} ensures consistency between the type maps $\Gamma_{\symb{fut}}$,
$\Gamma_{\symb{act}}$, and the overall security assignment {\it sec}.
It looks rather complex but it essentially only scoops up what has been prepared by
the other rules. The first two clauses ensure that the domains of activities coincide 
with the configuration domain and similarly for futures that the future type map 
$\Gamma_{\symb{fut}}$ is defined over precisely all futures
in all activities.  The third clause integrates the security specification 
$\symb{sec}$ to be 
respected by the individual security assignments of activities. 
The last large clause of {\sc Type Configuration} specifies 
first that the activity types assigned 
to activity references by $\Gamma_{\symb{act}}$ coincide with their active object types.
The second part of that clause addresses the future types in $\Gamma_{\symb{fut}}$. 
Note, that in the context of this clause we may assume 
$\alpha = \symb{futact}(f_k)$ by Property \ref{prop:futact}.
The clause ensures that the types assigned by $\Gamma_{\symb{fut}}$ coincide with the
ones assigned by $\Gamma_{\symb{act}}$ to their home activity.
Additionally, this final clause ensures that the request $Q(f_k)$ must have the type
assigned by the future map $\Gamma_{\symb{fut}}$ for this future $f_k$ with the $PC$ that
corresponds to the $PC$ assigned by the security assignment in the home activity.

\subsection{Running Example: Type System Checks Example}
\label{sec:extest}
For the sake of argument, we illustrate the application of the type system
with an inconsistent constraint on the assignment {\it sec} for the example in Section \ref{sec:ex}. 
The extended implementation as discussed in Section \ref{sec:ifex} contains the following changed ord function (we repeat the code here for convenience).

\begin{small}
\[
 \begin{array}{l}
    \beta \big[\varnothing, \\
    \quad [ \text{ord} = \varsigma(y)\ \text{if}\ \text{{\it this}.income}\mathord{/}10^3 \mathord{\ge} 1\  \text{then}\
            1\ \text{else}\ 0 ],\\
    \  \quad \text{income} = \ldots\\
    \quad \big]
 \end{array}
\]
\end{small}

If we specify income as private, this extended version of the running example may contain an implicit 
illegal information flow.
Any security assignment $\symb{sec}$ that fulfills the constraint must be fallacious
since the call $\beta_i$.ord in the manager object $\alpha$ reveals information 
about the confidential ($H$) value of income. 
The type system rejects any such $\symb{sec}$ since no consistent type can be inferred for the 
configuration in this case as we illustrate next. 
The failed type checking thus proves that for the extended configuration all specifications
would have to specify income $\mapsto L$ because the assumption income $\mapsto H$ was inconsistent.

The global classification is derived according to the visibility relation (Definition \ref{def:vis})
from the example's configuration as $\delta_{\beta_i} \sqsubseteq \da$ for all $i$. 
To be able to type the call to $\beta_i.\text{ord}$ in manager object $\alpha$
this method must be an $L$-method according to {\sc Type Call}. Hence, we need
to have the following extended constraint on {\it sec}.
\[ 
\begin{array}{ll}
\symb{sec} \supseteq & \{\text{ord}\mapsto L, 
 \text{income}\mapsto H\} 
\end{array} 
\]
The third clause of {\sc Type Configuration}, \ie 
$\bigcup_{\alpha \in \dom(C)} \symb{ass}(\Gamma_{\symb{act}}(\alpha)) \subseteq \symb{sec}$,
gives us the constraint 
$\symb{ass}(\Gamma_{\symb{act}}(\beta_i)) \supseteq \{ \text{ord}\mapsto L, \text{income}\mapsto H\}$
since {\it sec} and $\symb{ass}(\Gamma_{\symb{act}}(\beta_i))$ are both functions.

We show now that $\beta_i$ (for an arbitrary $i$ in the configuration) cannot be typed with this
type constraint.
The final step in a type inference to arrive at a type $\Gamma_{act}(\beta_i)$
for $\beta_i$ can only be an instance of {\sc Type Object} which looks as follows.
%
\begin{mathpar}
  \inferrule [Instance Type Object] 
    { \symb{this}\!:\! \Sigma_{\beta_i} \coloncolon []\!:\! \Sigma_{\beta_i}
      \coloncolon \varnothing; A(\text{ord}) \vdash  t_{\text{ord}}: \Sigma_{\beta_i} \\\\
      \symb{this}\!:\! \Sigma_{\beta_i} \coloncolon []\!:\! \Sigma_{\beta_i}
      \coloncolon \varnothing; A(\text{income}) \vdash  t_{\text{income}}: \Sigma_{\beta_i}}
    {\varnothing; \sqcup \{A(\text{ord}), A(\text{income})\} \vdash\\\\
      [\text{ord}  = \varsigma(y)\ t_{\text{ord}}, \text{income} = \varsigma(y)\ t_{\text{income}}]: \Sigma_{\beta_i}}
\end{mathpar}
%
We write $\Sigma_{\beta_i}$ for $\Gamma_{act}(\beta_i)$,
$A = \symb{sec}(\Sigma_{\beta_i})$, and 
$t_{\text{ord}} = \text{if}\ \text{{\it this}.income}\mathord{/}10^3 \mathord{\ge} 1\  \text{then}\ 1\ \text{else}\ 0$. 
In fact, a more technical definition of $t_{\text{ord}}$ is
\[\begin{array}{cl}
t_{\text{ord}} = & (((\symb{true}.\text{if} := (\symb{this}.>0(\symb{this}.\text{div}_{10^3}(\symb{this}.\text{income})))\\
& \ \ \ ).\text{then} := 1).\text{else:= 0}).\text{if}([]).
\end{array}
\]
where {\it true} is a boolean object containing methods if, then, and else.
The details of this boolean object and its typing as well as the details of the following
abridged reasoning are contained in Appendices $A$ and $B$.
The main point that we can see from this implementation is that the 
type $A(\text{ord})$ is coerced by the type $A(\text{if})$, \ie 
it must hold that $A(\text{ord}) = A(\text{if})$ in $\Sigma_{\beta_i}$.
This is the case, because $t_{\text{ord}}$ is a call to the method if. 
According to the rule {\sc Type Call}, the $PC$ must thus be $S_{\text{if}}$ 
which corresponds here to $A(\text{if})$ and coincides with the $PC$ 
$A(\text{ord})$ in the above instance of {\sc Type Object}, \ie $A(\text{ord}) = A(\text{if})$.
Now, the remaining argument just shows that $A(\text{if})$ must be $H$.
In short form, the reasoning for the latter goes as follows. 
By assumption, $A(\text{income})$ must be $H$. Thus
according to {\sc Type Call} and {\sc Val Self}, {\it this}.income is typeable 
only with $PC$ as $H$. We must apply {\sc Type Call} twice, to type
$\symb{this}.>0(\symb{this}.\text{div}_{10^3}(\symb{this}.\text{income}))$.
The $PC$ for typing this is $H$ each time because it must be the 
same as the $PC$ (named $S_j$) in typing the parameter (named $t$ in the 
rule {\sc Type Call}) and the previous typing of the parameter 
{\it this}.income has a $H$-$PC$.
The $PC$ in the application of the rule {\sc Type Update} is then also coerced
to $H$ in the typing of the newly inserted body method $l_j$, here ``if''. 
Hence, this update coerces the $PC$ $S_{\text{if}}$ to be $H$, \ie 
$A(\text{if})$ to be $H$. 
The two following updates do not change the security type of the method if.
We are finished since as we have seen above 
$A(\text{if}) = A(\text{ord})$. Thus, $A(\text{ord})$ must be $H$ and cannot be
typed $L$ as would be necessary to call this method remotely in $\alpha$.
The typing fails. We have a contradiction to the initially required specification 
that income be private. Since this was the only assumption, if follows by contraposition
that income must be $L$ to make the configuration typeable.

This illustrates the correctness of the type system by example: the configuration $C$ of 
our running example cannot be typed with the constraint income $\mapsto H$ since any 
attempt to infer a type $\langle \Gamma_{\symb{act}}, \Gamma_{\symb{fut}}, \symb{sec} \rangle$ 
for it fails.
The type inference reveals the dependency between ord and income: a security
leak because it would enable implicit information flows from $\beta_i$'s private part to
$\chi$.

In the following, we provide general proofs showing that the type system is sound, i.e., it
generally implies security not just for the example.

\section{Properties}
\label{sec:props}

\subsection{Preservation}
Type safety includes always a preservation theorem: if a program can be typed,
the type has to be preserved by the evaluation of the program -- otherwise the guarantees
encoded in the types would be lost. In our case, since configurations dynamically change during the
evaluation with the reduction relation $\dist$, the preservation has a slightly unusual form as the 
configuration type actually changes. But this change is conservative, \ie dynamically
created new elements are assigned new types but old types persist, as represented below 
by $\subseteq$.
Alongside the configuration types, also the security class lattice is extended likewise
in a conservative way by extension of the visibility relation.

\begin{theorem}[Preservation]
\label{thm:subred}
\begin{small}
\[
\left( 
\begin{array}{l}
\! \vdash C\!:\!\langle \Gamma_{\symb{act}}, \Gamma_{\symb{fut}}, \symb{sec}\rangle\! \\ C\dist C'
\end{array}
\!\right)\!
\Rightarrow \exists\,
\Gamma_{\symb{act}}', \Gamma_{\symb{fut}}'.\, \vdash  C'\!:\! \langle \Gamma_{\symb{act}}', \Gamma_{\symb{fut}}', \symb{sec} \rangle
\]
where $\Gamma_{\symb{act}}\subseteq\Gamma_{\symb{act}}'$ and $\Gamma_{\symb{fut}}\subseteq\Gamma_{\symb{fut}}'$.
\end{small}
\end{theorem}

The proof of this theorem has two parts. The first part shows a local preservation property
for the part of the type system that describes secure method calls at the level of objects, i.e.,
the rules depicted in Tables \ref{tab:sectypeord} and \ref{tab:sectype}. The second part of the proof
addresses the typing rules at the global level, i.e., the configuration typing rules depicted in 
Table \ref{tab:typeconf}. Both proofs are straightforward using the induction schemes corresponding to
the inductive rule definitions of the type rule definitions. Albeit the relatively small size of the 
computation model \aspfunp, these rules are fairly complex. Hence to avoid mistakes in these proofs we have 
formalized them in Isabelle/HOL. 
The Isabelle/HOL sources can be found at \url{https://sites.google.com/site/floriankammueller/home/resources}.

\subsection{Confinement}
\label{sec:conf}
Confinement is the property of our type system encoding the principal idea of
the security model: if a method of an object can be called remotely, it must be a
public $L$ method.
As a preparation to proving confinement, we present next a chain of lemmas that lead up to it.
Let $o$ be an object and $T$ be an arbitrary type environment throughout the following formal
statements.


The type rules for subsumption allow that 
types of objects can be ``lifted'', \ie
objects can have more than one type. 
We lose uniqueness of type judgments. To overcome this, we use a well-known trick (already
been used in the Hindley-Milner type system for ML to accommodate polymorphic types) 
to regain some kind of uniqueness: minimal types. 
\begin{definition}[Minimal Type]
Define the minimal type in the $PC$ context of $S  \in \{L, H\}$ as follows.
\[T; S \vdash_{{\rm \footnotesize ML}} o: (A, \delta) \equiv 
  \left\{
   \begin{array}{l}
  T; S \vdash o: (A, \delta)\, \wedge  \\
    \forall S',A', \delta'. \\
    T; S' \vdash o: (A', \delta') \Rightarrow 
     \left(
     \begin{array}{c}
        S \leq S' \\
        \delta \sqsubseteq \delta' \\
        A \subseteq A'
     \end{array}
     \right)
   \end{array}\right.
\]
\end{definition}
This provides at least that minimal types of local typings are unique.
\begin{lemma}[Minimal Type Uniqueness]
Let $S \in \{L, H\}$.
If $T; S \vdash_{{\rm \footnotesize ML}} o: (A, \delta)$ and $T; S \vdash_{{\rm \footnotesize ML}} o: (A', \delta')$, 
then $\delta = \delta'$.
\end{lemma}

A slightly stronger form of that previous lemma exists for $H$ $PC$s.
\begin{lemma}[High PC Uniqueness]
If $T; H \vdash o: (A, \delta)$ and $T; H \vdash_{{\rm \footnotesize ML}} o: (A, \delta')$, then $\delta = \delta'$.
\end{lemma}

Using slight generalization and contraposition, the previous lemma can be strengthened to the following key
lemma for confinement.
\begin{lemma}[Abstract Confinement]
\label{lem:absconf}
If $T; S \vdash o: (A, \delta)$ and $\delta_0 \sqsubset \delta$ and $T; S_0 \vdash_{{\rm \footnotesize ML}} o: (A, \delta_0)$, 
then $S_0 = L$.
\end{lemma}
The following key fact, about the minimal type for futures provides the anchor to apply 
Abstract Confinement and arrive at Confinement.
\begin{proposition}[Minimal Future Type]
\label{prop:minfut}
Let $\vdash C: \langle \Gamma_{\symb{act}}, \Gamma_{\symb{fut}}, \symb{sec} \rangle$,
$f_k \in \dom(\Gamma_{\symb{fut}})$, and $\alpha = \symb{futact}(f_k)$ the home activity of $f_k$. Then 
\[  
\langle \Gamma_{\symb{act}},\Gamma_{\symb{fut}}, \symb{sec}\rangle, A_\alpha(\symb{futlab}(f_k)) \vdash_{{\rm \footnotesize ML}} 
f_k: \Gamma_{\symb{act}}(\alpha) \,.
\]
\end{proposition}

\begin{theorem}[Confinement]
\label{thm:conf}
If a future $f_k$ is typeable with an arbitrary $PC$ $S$ as of type $\delta$ strictly larger than
the global level of $f_k$'s home activity $\alpha$, then $f_k$ has been initially generated from a call
to an $L$ method of $\alpha$.
Formally, 
let $\vdash C: \langle \Gamma_{\symb{act}} \cup \symb{sec}, \Gamma_{\symb{fut}}\rangle$, $\alpha[Q,a] \in C$, 
and $f_k \in \dom(Q)$ with
\[\langle \Gamma_{\symb{act}}, \Gamma_{\symb{fut}}, \symb{sec} \rangle, T; S \vdash f_k: (A_x,\delta)
\ {\rm where}\ \Gamma_{\symb{act}}(\alpha) \sqsubset \delta \,.\] 
Then 
\[A_\alpha(\symb{futlab}(f_k)) = L\,.\]
\end{theorem}

The proof of confinement is basically just a combination of Lemma \ref{lem:absconf}
and Proposition \ref{prop:minfut}. The chain of lemmas and confinement have been proved in Isabelle/HOL
as well.

\subsection{Noninterference}
Confinement can be considered as a simple security property because it 
is similar to a safety property:
confinement is preserved on every trace of execution of a configuration. 
Intuitively it seems to imply confidentiality of private parts but this is only true 
for direct information flows.
Confidentiality 
necessitates that no information is leaked to an outsider 
even considering implicit information flows as 
described in Section \ref{sec:sec}. Based on those observations, we define the general property 
of confidentiality as {\it noninterference}, informally meaning that an attacker cannot learn anything despite
his ability to observe configurations on all runs while comparing values that he can see: a difference
in the value of the same call allows deductions about a change in hidden parts. The formal 
definition of noninterference for active objects in general \cite{kam:12}
is a bisimulation over the indistinguishability 
relation $\inda$ on configurations. We omit the rather technical definition of indistinguishability 
referring to Appendix D. Essentially, indistinguishability says that $C$ and $C_1$ appear equal to the 
attacker $\alpha$'s viewpoint  
even if they differ in secret parts; noninterference means that this appearance is preserved by the evaluation of configurations. 
\begin{definition}[$\alpha$-Noninterference]
\label{def:ni}
If configuration $C$ is indistinguishable to any $C_1$ for $\alpha$ with respect to {\it sec} 
and remains so under the evaluation of configurations $\dist$, 
then $C$ is $\alpha$-noninterfering.
Formally, we define $\alpha$-noninterference $C$ {\it sec} as follows.
\[ \left(\begin{array}{c} C \dist C' \\ 
                      C \inda C_1  
         \end{array}\right)
\imp 
\exists\ C_1'.\left( \begin{array}{c} C_1 \diststar C_1' \\
                        C' \inda C_1'
       \end{array}\right)
\]
\end{definition}
A main result for our security type system is {\it soundness}: a well-typed configuration is secure; 
$\alpha$-noninterference holds for the configuration, i.e., it does not leak information.
\begin{theorem}[Soundness]
\label{thm:ni}
For any well-typed configuration $C$, 
we have noninterference with respect to $\alpha \in C$, \ie
\[ 
\left(
\begin{array}{c}
\vdash C: \langle \Gamma_{\symb{act}}, \Gamma_{\symb{fut}}, \symb{sec}\rangle \\
 \alpha[Q,a] \in C 
\end{array}\right)
\Rightarrow 
 \alpha{\mbox{\rm {-noninterference}}}\ C\ \symb{sec} \,.
\]

\end{theorem}
The proof of this theorem is a case analysis distinguishing the 
cases where a reduction step of the configuration has happened 
in the $\alpha$-visible part or outside it. 
In the latter case, a difference in the visible part would mean a breach of confinement.
Within the visible part, a straightforward case analysis shows that what is possible
in one configuration must also be possible in the other, indistinguishable, one, since 
those parts are isomorphic; hence the same reduction rules apply.
We have formalized the definitions of indistinguishability, noninterference, and 
multilateral security, as well as the statements of the theorems in Isabelle/HOL
-- only the soundness proof is not yet formalized but a detailed paper proof is
contained in Appendix E.

The parameterization of the attacker as an active object $\alpha$ grants the 
possibility to adapt the noninterference predicate.
If we universally quantify $\alpha$ in our definition of noninterference, we 
obtain a predicate where each object could be the attacker 
corresponding to multi-lateral security.
\begin{definition}[Multi-Lateral Security]
If a configuration $C$ is $\alpha$-noninterfering for all
$\alpha \in \dom(C)$ then multi-lateral security holds for $C$.
\end{definition}
Since no $\alpha$ is fixed in the type statement, the soundness theorem holds
for any $\alpha$ if the configuration is well-typed. Hence, well-typing implies
immediately multi-lateral security.

\section{Related Work and Conclusions}
\label{sec:concl}
The main difference of our approach is that we specifically address functional active 
objects. We also use a non-standard security model \cite{kam:12} for 
multi-lateral security tailored to distributed active objects.
Other work on actor security, e.g. \cite{hmss:07}, is based on message passing
models different to our high level language model. The paper \cite{achl:07}
addresses only direct information flows in active objects.
The priority program Reliably Secure Software Systems (RS3) of the 
German Research Foundation (DFG) \cite{rs3:10} addresses in its part project MoVeSPAcI 
\cite{pf:11} security of actor systems using an event based approach without futures.

The Distributed Information Flow Control (DIFC) approach \cite{ml:97} provides
support for Java programs (Jif) to annotate programs with labels ``Alice'' 
and ``Bob'' for information flow control. In this approach objects are not first class
citizen. The formal model \cite{zm:07} uses a lambda calculus $\lambda_{\symb{DSec}}$ to 
accommodate the rich hierarchy of labels but (Java) objects are not in the calculus.
They use an elegant approach to prove noninterference of a type system for labels
pioneered by \cite{ps:03} Pottier and Simonet. This approach does not apply to parallel 
languages since the evaluation order of parallel processes is not deterministic.

Reactive Noninterference, e.g. \cite{DBLP:journals/jlp/MatosBC07}, \cite{bpswz:09},
adopts a reactive system view. Some of these works, 
e.g. \cite{occ:06}, use a while language in their formal models and bisimulation
based noninterference notions but the semantics is message passing by events.

The language based approach has its 
beginnings in \cite{vsi:96,vs:97}.
\cite{sv:98} offer the first model of language based 
information flow control  for concurrency, later refined by Boudol and Castellani
addressing scheduling problems and related timing leaks.
Many works have followed this methodology (see \cite{sm:03} for an overview).
However, most works consider imperative while languages with various extensions 
like multi threading. 
Barthe and Serpette  \cite{bs:00} have considered security type systems for $\varsigma$-objects
but no distribution. Later, Barthe and others 
\cite{bpr:07,kam:08} provide information flow control for Java-like languages. 
Sabelfeld and Mantel consider message passing in distributed programs \cite{sm:02,ms:03}.
These works use the secure channel abstraction, i.e. connecting remote processes of
the same security class via secure channels integrating security primitives.

Distributed security has also been considered in many works in the setting of process algebras
most prominently using pi calculus by \cite{Mil:89}
(see \cite{fg:95} and \cite{rsglr:00} providing overviews). 
Commonalities of process algebra based security to our work are the bisimulation notion of 
noninterference and asynchronous communication. 
The spi calculus by \cite{ag:97} extends the pi calculus with constructs for encryption and
decryption. It is thus a forerunner for current work that integrates encryption primitives
into languages, most prominently homomorphic encryption \cite{fpr:11}.
The applied pi calculus \cite{af:01} in contrast is a generalisation of Milner's original
pi calculus with equational theories, i.e. functions and equations.
Thereby, extensions by cryptographic primitives are possible.
The applied pi calculus is used for security protocol verification. An 
implementation is the model checker ProVerif by \cite{cb:13}. 
There is a line of research on mobile calculi that use purely functional concurrent
calculi. A few representative papers are by \cite{hvy:00} on the pi calculus and
\cite{hr:02} for the security pi calculus. 
An impressive approach on information flows for distributed languages with mobility and states 
\cite{mc:11} first introduces declassification. Similar work is by \cite{bcc:01} also studying 
noninterference for distribution and mobility for Boxed Ambients.
\cite{bf:08} deviates from the applied pi calculus generality focusing on core abstractions
for security in distributed systems, like secure channels \cite{bf:10}.
In common with these works are modeling distributed system by a calculus but 
none of the pi calculus related work focuses on active objects while we do not
consider cryptographic primitives.
An interesting perspective would be to investigate the relationship between confinement
and effects of cryptographic primitives.
We also use a bisimulation-based equivalence relation to express noninterference.
In the applied pi calculus, for example, the notion of a static equivalence, similar
to our indistinguishability is used in addition to observational equivalence
that corresponds to our notion of noninterference (see e.g. \cite{dkr:10}). 

This work presented a formal framework for the security of active objects based on 
the semi-lattice security model that propagates confinement.  
We presented a safe security type system, 
that verifies the confinement property  and
is sound, i.e., checks security, with respect to a dedicated formal notion of noninterference, 
or more generally, multi-lateral security. 
\aspfun makes secure down-calls possible and is still applicable bi-directionally as 
illustrated by implementing the NSPK protocol. 
The proofs have been in large parts 
formalized 
in Isabelle/HOL.
An implementation of functional active objects is given by Erlang Active Objects in \cite{fk:10}
also providing a simple extension by a run-time monitor for confinement \cite{fk:11}.

\bibliographystyle{IEEEtranS}
\bibliography{biblio}

\section*{Appendix}
\subsection*{A: Booleans and conditional in the $\varsigma$-calculus and their security types}
To prepare for the type inference, we need the implementation
of the boolean datatype and the {\it if-then-else} 
in the $\varsigma$-calculus, i.e. in the local calculus of \aspfunp. 

\[
\begin{array}{l}
\text{true} = \left[\,\text{if} = \varsigma(y) \symb{this}.\text{then}(y), \right. \\
    \qquad\qquad \left. \text{then} = \varsigma(y) [], \text{else} = \varsigma(y) []\,\right]\\
\text{false} = \left[\,\text{if} = \varsigma(y) \symb{this}.\text{else}(y), \right.\\
    \qquad\qquad \left. \text{then} = \varsigma(y) [], \text{else} = \varsigma(y) []\,\right]\\
\text{if}\ b\ \text{then}\ c\ \text{else}\ d = \\
     \qquad\qquad((b.\text{then} := \varsigma(y) c).\text{else} := \varsigma(y) d).\text{if}([])
\end{array}
\]

In the third line above, $\symb{this},y \notin FV(c) \cup FV(d)$; $[]$ denotes the empty
object. The definition shows how -- similar to $\lambda$-calculus -- the functionality of
the constructor is encoded in the elements of the datatype: when $b$ is true 
its method if delegates to the method then, filled with
term $c$, when false, if delegates to else, executing term $d$.

Typing of the {\it if-then-else} construct is a base test for an information flow
type system as this construct is the basic example that gives rise to implicit
information flows. We will thus here illustrate how the security type rules presented
in this paper establish that the guard of the {\it if-then-else} construct, the if, must
be typed with the same $PC$ as the branches, \ie then and else. Then it immediately follows
that if the method if has $H$-$PC$ then the branches must have $H$-$PC$ as well.
The reasoning instantiates type rules showing the constraints that follow for the
security assignment in the security type $\Sigma_{\text{ifte}}$.

A condition $b$ in the method if of an {\it if-then-else} object evaluates 
to either $true$ or $false$. We consider those two possibilities and infer
their types and the resulting constraints. 

To type {\it true}, we initially type {\it this} which can be done only by rule {\sc Val Self}
leading to the following typing where $\Sigma_{\text{ifte}} = (A_{\text{ifte}},\delta_{\text{ifte}})$ 
is the security type for the {\it if-then-else} object and 
$M_{\text{ifte}} = \sqcup  \{A_{\text{ifte}}(\text{if}), A_{\text{ifte}}(\text{then}), A_{\text{ifte}}(\text{else}) \}$.
We use the arbitrary set of additional type assumptions $T$ provided by the rule to integrate
the type assumption for $y$ already here. It is needed further down for typing the object but only
formally.

\begin{small}
\begin{mathpar}
\inferrule[Instance Val Self]
{this\!:\! \Sigma_{\text{ifte}} \coloncolon(y\!:\! \Sigma_{\text{ifte}}) \coloncolon T; M_{\text{ifte}} \vdash \symb{this}\!:\! \Sigma_{\text{ifte}}}
{}
\end{mathpar}
\end{small}

We then apply the rule {\sc Type Call} to infer a type for {\it this}.then.
The following instance of that rule sets the parameters such that it can be applied to the
previous {\sc Instance Val Self}.

\begin{small}
\begin{mathpar}
\inferrule [Instance Type Call]
{\symb{this}\!:\! \Sigma_{\text{ifte}}\coloncolon y\!:\! \Sigma_{\text{ifte}} \coloncolon T; M_{\text{ifte}} \vdash \symb{this}\!:\! \Sigma_{\text{ifte}}\\\\
\symb{this}\!:\! \Sigma_{\text{ifte}} \coloncolon y\!:\! \Sigma_{\text{ifte}} \coloncolon T; A_{\text{ifte}}(\text{then})\vdash []\!:\! \Sigma_{\text{ifte}}
}{\symb{this}\!:\! \Sigma_{\text{ifte}} \coloncolon y\!:\! \Sigma_{\text{ifte}} \coloncolon T; A_{\text{ifte}}(\text{then}) \vdash \symb{this}.\text{then}([])\!:\! \Sigma_{\text{ifte}}}
\end{mathpar}
\end{small}

Now, to type the {\it true} object including its fields then and else
we next need an instance of {\sc Type Object}.

\begin{small}
\begin{mathpar}
  \inferrule [Instance Type Object] 
  { \symb{this}\!:\! \Sigma_{\text{ifte}} \coloncolon y\!:\! \Sigma_{\text{ifte}} \coloncolon T; 
                  A_{\text{ifte}}(\text{if})\vdash \symb{this}.\text{then}([]) :\Sigma_{\text{ifte}}\\
     \symb{this}\!:\! \Sigma_{\text{ifte}} \coloncolon y\!:\! \Sigma_{\text{ifte}} \coloncolon T; 
                  A_{\text{ifte}}(\text{then})\vdash [] :\Sigma_{\text{ifte}}\\
      \symb{this}\!:\! \Sigma_{\text{ifte}} \coloncolon y\!:\! \Sigma_{\text{ifte}} \coloncolon T;
                  A_{\text{ifte}}(\text{else})\vdash [] :\Sigma_{\text{ifte}}\\
     }
    {T; M_{\text{ifte}} \vdash \text{true} = \left[\,\text{if} = \varsigma(y) \symb{this}.\text{then}(y),
                  \text{then} = \varsigma(y) [], \text{else} = \varsigma(y) []\,\right] : \Sigma }
\end{mathpar}
\end{small}

The main observation is that the following constraint must hold for $A_{\text{ifte}}$
\[ A_{\text{ifte}}(\text{if}) = A_{\text{ifte}}(\text{then}) \]
because this is necessary for the first proviso of the 
above instance to be matched with the previous type derivation for 
$\symb{this}.\text{then}([])$ by {\sc Instance Type Call}.

With a very similar argument for typing {\it false}, \ie $\symb{this}.\text{else}([])$, 
we arrive at a similar constraint.
\[ A_{\text{ifte}}(\text{if}) = A_{\text{ifte}}(\text{else}) \]

Since for an arbitrary {\it if-then-else} guard $b$ we have to allow both values
{\it true} and {\it false} as possible outcome we have to combine the constraints and
conclude for $A_{\text{ifte}}$ the following overall constraint.
\[ 
A_{\text{ifte}}(\text{if}) = A_{\text{ifte}}(\text{then}) = A_{\text{ifte}}(\text{else}) 
\]
The update of the methods then and else does not change the $PC$ and thus
preserves the security assignment and the constraints.
This constraint is what we expect for information flow security. If the guard of an
{\it if-then-else} can only be typed in a $H$-$PC$ then its branches must also be
``lifted'' to $H$. Only if the guard can be typed in a $L$-$PC$, can the branches also 
be typed in $L$-$PC$.

\subsection*{Note on typing constants}
In the above type rule instances we have used typings for constants, for example,
the empty object $[]$ as granted and did not refine them any further.
\[
\symb{this}\!:\! \Sigma_{\text{ifte}} \coloncolon y\!:\! \Sigma_{\text{ifte}} \coloncolon T; A_{\text{ifte}}(\text{else})\vdash [] :\Sigma_{\text{ifte}}
\]
A word is in order to explain how these are constructed and their types are derived. 
A simple way to integrate the empty object into an activity is to add a method empty and then 
replace all $[]$ by {\it this}.empty. The security assignment should be $A(\text{empty}) = L$. 
Then, we can use {\sc Type Call} to have $\dots L \vdash {\it this}.\text{empty}: \Sigma_{\text{ifte}}$ 
and from there derive the above 
$\dots;A_{\text{ifte}}(\text{else})\vdash {\it this}.\text{empty}: \Sigma_{\text{ifte}}$.
However, $[]$ (and other commonly used plain objects) can more practically be considered as
{\it activities without any $H$ methods} that are included as a ``data base''
in a configuration. Then, an occurrence of $[]$ is literally the activity named ``empty object'',
\ie $[]$ is an activity reference. 
For the typing, the natural type of the empty object is given as the empty security assignment 
$\varnothing$, \ie the partial function that is undefined for all inputs, and the
bottom element $\bot$ of the visibility semi-lattice which corresponds to the empty set of
activity names.
\[
\langle \Gamma_{\symb{act}}, \Gamma_{\symb{fut}}, \symb{sec} \rangle; L \vdash []: (\varnothing, \bot)
\]
By definition this typing with $L$-$PC$ as $(\varnothing, \bot)$ for the empty object enables
typing $[]$ ``into'' any other activity type $(A, \delta)$ because $\varnothing \subseteq A$ and
$\bot \sqsubseteq \delta$. Thus -- by {\sc SecAss Subsumption}  and {\sc Glob Subsumption} -- 
$\dots; L \vdash [] : (A, \delta)$. 

A similar type and subtyping argumentation goes for other constants, for example $0$ or $1$, used in
the running example. Similar to Church numerals simple term representation can be given to them
in \aspfunp. Such constant activities $\eta$ must have their methods all assigned to $L$, i.e.,
their security assignment $A$ maps all method names of $\eta$ to $L$. Then the
$PC$ of the activity $\eta$ is also $L$ because it is given as $\sqcup \{L\}$ according to 
the rule {\sc Type Configuration}. The global level of a constant activity like $\eta$ 
is defined as the set $\{eta\}$. If the constant $\eta$ is used by referencing
it in other activities of the configuration, the name $\eta$ becomes part of the
other activities' global levels.

\subsection*{B: Running Example -- Details on Typing}
The following shows why the example configuration presented as running example cannot be typed
with income $\mapsto H \in \symb{sec}$. 

\subsection*{Implementation}
The quicksort function  is described in Section \ref{sec:ex}.
The manager activity that controls the ordering of a list and the sorting object $\chi$ that
calls the ord method in $\beta$-objects are repeated here for convenience of the reader.

\noindent%
\begin{small}
\[
 \begin{array}{ll}
   & \alpha \big[\varnothing, [ \text{manage} = \varsigma(y) \text{{\it this}.sort({\it this}.list)}, \\
  & \  \qquad\ \text{sort} = \varsigma(y)\ \chi.\text{qsort}(y),\\
  & \  \qquad\ \text{list} = \ldots ] \big]
 \end{array}
\]
\end{small}

\noindent%
\begin{small}
\[
 \begin{array}{l}
   \chi \big[\varnothing, \\
   \quad [ \text{qsort} = \varsigma(y)\ \text{if}\ y = []\ \text{then}\ [] \\
   \ \qquad \qquad \qquad \ \ \text{else let}\ (a::l) = y \\
   \ \qquad \qquad \qquad \qquad \qquad (l_1,l_2) = \text{{\it this}.part}\ (a, l) \\
   \ \qquad \qquad \qquad \qquad \qquad\ l_1' = \text{if}\ \#l_1 \mathord{\leq} 1\ \text{then}\ l_1\ \text{else {\it this}.qsort}(l_1) \\
   \ \qquad \qquad \qquad \qquad \qquad\ l_2' = \text{if}\ \#l_2 \mathord{\leq} 1\ \text{then}\ l_2\ \text{else {\it this}.qsort}(l_2) \\
   \ \qquad \qquad \qquad \qquad \ \ \text{in}\ l_1' @ [a] @ l_2'\ \\
   \ \qquad \qquad \qquad \ \ \text{end},\\
   \ \quad \text{part} = \varsigma(p,y)\ \text{if}\ y = []\ \text{then}\ ([],[])\\
   \ \qquad \qquad \qquad \quad\, \text{else let}\ (a\mathord{::}l) = y \\
   \ \qquad \qquad \qquad \qquad \qquad \ \, (l_1,l_2) = \text{{\it this}.part}\ (p,l) \\
   \ \qquad \qquad \qquad \qquad \quad \ \text{in if}\ p \mathord{<} a.\text{ord then}\ (l_1, a\mathord{::}l_2)\ 
                                             \text{else}\ (a\mathord{::}l_1,l_2)\ \\
   \  \qquad \qquad \qquad \quad\, \text{end}\\
   \quad ]\big]
  \end{array}
\]
\end{small}

The extended method ord that bears a dependency between ord and income,
\begin{small}
\[
 \begin{array}{l}
    \beta \big[\varnothing, \\
    \quad [ \text{ord} = \varsigma(y)\ \text{if}\ \text{{\it this}.income}\mathord{/}10^3 \mathord{\ge} 1\  \text{then}\
            1\ \text{else}\ 0 ],\\
    \  \quad \text{income} = \ldots\\
    \quad \big]
 \end{array}
\]
\end{small}
is not typeable for any security assignment $\symb{sec}$ that imposes the constraint that
method income $\mapsto H$. 
The following type inference elaborates that the type system rejects any
security assignment that contains the constraint income $\mapsto H$. 
It illustrates how the security assignment $A_{\beta_i} = \symb{ass}(\Gamma_{\symb{act}}(\beta_i))$ is inferred.

\subsection*{Typing remote call implies ord $\mapsto L$}
Since the method ord is called remotely in $\chi$ via $\alpha$ we need that ord $\mapsto L$,
which cannot be possible because of the dependency in the above implementation.
To be able to type the call to $\beta_i.\text{ord}$ in the object  $\chi$ 
this method must be an $L$-method according to {\sc Type Call} and {\sc Glob Subsumption}.
More precisely, let $(A_{\beta_i}, \delta_{\beta_i}) = \Gamma_{\symb{act}}(\beta_i)$.
We have that 
$\langle \Gamma_{\symb{act}}, \Gamma_{\symb{fut}}, \symb{sec} \rangle, \varnothing; M_{\beta_i} \vdash \beta_i: (A_{\beta_i}, \delta_{\beta_i})$ because of {\sc Type Active Object Reference} and $\beta_i \in \dom(C)$.
The $PC$ is $M_{\beta_i} = \sqcup_{j \in \dom{A_{\beta_i}}} A_{\beta_i}(j)$ where $A_{\beta_i}$ needs to be inferred 
in the process. We can use next {\sc Type Call} to type 
$\langle \Gamma_{\symb{act}}, \Gamma_{\symb{fut}}, \symb{sec} \rangle, \varnothing; A_{\beta_i}(\text{ord}) \vdash \beta_i.\text{ord}: (A_{\beta_i}, \delta_{\beta_i})$.
However, to type $\beta_i$.ord in the context of the object $\chi$ it needs to be typed as
$(A_{\beta_i}, \delta_{\chi})$ with global type component $\delta_\chi$. This upgrading of the
call can only be achieved by application of rule {\sc Glob Subsumption} which requires
that $\delta_{\beta_i} \sqsubseteq \delta_\chi$ which is true but also requires that the $PC$
of the typing $\langle \Gamma_{\symb{act}}, \Gamma_{\symb{fut}}, \symb{sec} \rangle, \varnothing; A_{\beta_i}(\text{ord}) \vdash \beta_i.\text{ord}: (A_{\beta_i}, \delta_{\beta_i})$, \ie $A_{\beta_i}(\text{ord})$, is $L$.

\subsection*{Typing $\beta_i.\text{ord}$ at global level $\delta_{\beta_i}$ only with $H$-$PC$}
The next part of the argument states that the only type that can be inferred for a call $\beta_i.\text{ord}$
is $T; H \vdash \beta_i.\text{ord}: (A_{\beta_i}, \delta_{\beta_i})$, \ie with $H$-$PC$.
This is because types for calls can only be inferred by rule {\sc Call} and $A_{\beta_i}(\text{ord}) = H$ 
which coerces the $PC$ according to rule {\sc Call} to $H$. 
For clarity of the exposition, we omit in the following 
$\langle \Gamma_{\symb{act}}, \Gamma_{\symb{fut}}, \symb{sec} \rangle$ in front of the typings.
Since we want to arrive at $\vdash C : \langle \Gamma_{\symb{act}}, \Gamma_{\symb{fut}}, \symb{sec} \rangle $,
by the inversion principle of inductive type definitions,
all provisos of {\sc Type Configuration} have to be true.
Since
\[\beta_i[\varnothing, [\text{ord}  = \varsigma(y)\ t_{\text{ord}}, \text{income} = \varsigma(y)\ t_{\text{income}}] ] \in C, \]
the fourth proviso, first clause, of {\sc Type Configuration} yields
\[\varnothing, M_{\beta_i} \vdash [\text{ord}  = \varsigma(y)\ t_{\text{ord}}, \text{income} = \varsigma(y)\ t_{\text{income}}]: (A_{\beta_i}, \delta_{\beta_i}).\]

The coercion of $A_{\beta_i}(\text{ord})$ to $H$
is a consequence of the typing of the object $\beta_i$ with an instance of rule {\sc Type Object}.

\begin{small}
\begin{mathpar}
  \inferrule [Instance Type Object] 
    { \symb{this}\!:\! \Sigma_{\beta_i} \coloncolon []\!:\! \Sigma_{\beta_i}
      \coloncolon \varnothing; A_{\beta_i}(\text{ord}) \vdash  t_{\text{ord}}: \Sigma_{\beta_i} \\\\
      \symb{this}\!:\! \Sigma_{\beta_i} \coloncolon []\!:\! \Sigma_{\beta_i}
      \coloncolon \varnothing; A_{\beta_i}(\text{income}) \vdash  t_{\text{income}}: \Sigma_{\beta_i}}
    {\varnothing; M_{\beta_i}
     \vdash [\text{ord}  = \varsigma(y)\ t_{\text{ord}}, \text{income} = \varsigma(y)\ t_{\text{income}}]: \Sigma_{\beta_i}}
\end{mathpar}
\end{small}

This instance enforces $A(\text{ord})$ to be the same as the $PC$ in the typing of
\[ 
\begin{array}{ccl}
t_{\text{ord}} & = & 
     \text{if}\ \text{{\it this}.income}\mathord{/}10^3 \mathord{\ge} 1\  \text{then}\ 1\ \text{else}\ 0 \\
   & = & (((\symb{true}.\text{if} := (\symb{this}.>0(\symb{this}.\text{div}_{10^3}(\symb{this}.\text{income})))\\
& & \ \ \ ).\text{then} := 1).\text{else:= 0}).\text{if}([]).
\end{array}
\] 
The only way to arrive at a type for $t_{\text{ord}}$ is by an application of {\sc Type Call}
as in the following instance.

\begin{small}
\begin{mathpar}
 \inferrule [Instance Type Call] 
    {  T; S \vdash  (((\symb{true}.\text{if} := (\symb{this}.>0(\symb{this}.\text{div}_{10^3}(\symb{this}.\text{income}))))\\\\
  .\text{then} := 1).\text{else:= 0}): \Sigma \\\\ 
       T; A_{\beta_i}(\text{if})  \vdash []: \Sigma}
    {T; A_{\beta_i}(\text{if}) \vdash t_{\text{ord}} : \Sigma}
\end{mathpar}
\end{small}

In order to match the conclusion of the above with the first proviso of the earlier
{\sc Instance Type Object}, the security assignment of ord is coerced to that of if
\[ A_{\beta_i} (\text{ord}) = A_{\beta_i} (\text{if}).
\]
We only need to show that $A_{\beta_i} (\text{if}) = H$ and we are finished.

\subsection*{Typing implies that $A_{\beta_i}(\text{if}) = H$}
The following chain of steps shows how a type for the body of ord and thus $A_{\beta_i}(\text{ord})$
must be inferred detailing how the security assignment parameter $A_{\beta_i}(\text{if})$ needs to 
be instantiated to $H$.
The chain of reasoning starts from the one specified security assignment 
income $\mapsto H$ in $\symb{sec}$ and shows that then also ord $\mapsto H$
which contradicts the above ord $\mapsto L$. Hence, no type can exist with
the constraint income $\mapsto H$ for this configuration.

$A_{\beta_i}(\text{income})$ is $H$ by constraint on $sec$ and thus $A_{\beta_i}$.
According to {\sc Val Self} with 
\[ M = \sqcup\{A_{\beta_i}(\text{income}), A_{\beta_i}(\text{ord}), \ldots\} = H\]
we get the following typing for {\it this}.
\[
\begin{array}{l}
T; H \vdash \symb{this}: (A_{\beta_i}, \delta_{\beta_i})
\end{array}
\]

According to {\sc Type Call}, {\it this}.income is typeable  
only with $PC$ as $H$ since $\{ \text{income} \mapsto H\} \subseteq A_{\beta_i}$.

\[
\begin{array}{l}
T; H \vdash \symb{this}.\text{income}: (A_{\beta_i}, \delta_{\beta_i})
\end{array}
\]

The previous typing feeds into rule {\sc Type Call} again but this time
for the parameter $t$. Since the $PC$ $S_j$ matches with $H$ we get again
a $H$-$PC$ coercing the method $\text{div}_{10^3}$ also to be assigned to $H$
in $A_{\beta_i}$.

\[
\begin{array}{l}
T; H \vdash \symb{this}.\text{div}_{10^3}(\symb{this}.\text{income}): (A_{\beta_i}, \delta_{\beta_i})
\end{array}
\]

In the same fashion, the previous considered as a parameter typing
{\sc Type Call} consequently coerces $A_{\beta_i}(>0)$ also to $H$:

\[
\begin{array}{l}
T; H \vdash \symb{this}.>0(\symb{this}.\text{div}_{10^3}(\symb{this}.\text{income})): (A_{\beta_i}, \delta_{\beta_i})
\end{array}
\]

We instantiate {\sc Update} as follows.

\begin{small}
\begin{mathpar}
 \inferrule [Instance Type Update]
    {\varnothing; S \vdash \text{true}: (A_{\beta_i}, \delta_{\beta_i}) \\\\
    \symb{this}\!:\! \Sigma_{\beta_i}\coloncolon []\!:\! \Sigma_{\beta_i} \coloncolon \varnothing; A_{\beta_i}(\text{if}) \vdash \\\\
     \symb{this}.>0(\symb{this}.\text{div}_{10^3}(\symb{this}.\text{income})): (A_{\beta_i}, \delta_{\beta_i})  }
    {\varnothing; \sqcup \{A_{\beta_i}(\text{if}), \dots \} ) \vdash \\\\
\text{true}.\text{if}[] := \symb{this}.>0(\symb{this}.\text{div}_{10^3}(\symb{this}.\text{income})): (A_{\beta_i}, \delta_{\beta_i}) }
\end{mathpar}
\end{small}

The first proviso, the typing for {\it true} can be inferred as shown in the previous section,
using rule {\sc Sec Ass Subsumption} in addition to embed it into $\beta_i$.
We spell out some portion of $M_{\beta_i} = \sqcup \{A_{\beta_i}(\text{if}), \dots \}$ above to
emphasize that $A_{\beta_i}(\text{if})$  is part of the $PC$. 
The dots stand for $A_{\beta_i}(\text{ord})$
and $A_{\beta_i}(\text{income})$ etc. To match the previous derivation above of
$\varnothing; H \vdash \symb{this}.>0(\symb{this}.\text{div}_{10^3}(\symb{this}.\text{income})) : (A_{\beta_i}, \delta_{\beta_i})$ to the second proviso in the above instance it is necessary to coerce
\[ A_{\beta_i}(\text{if}) = H. \]
We are finished here already because we have already shown above that $A_{\beta_i}(\text{if}) = A_{\beta_i}(\text{ord})$ which thus is $H$ contradicting the earlier requirement to be $L$.

For completeness, we continue the derivation of the body of $t_{\text{ord}}$. 
From the previous step above, we get the conclusion 

\begin{small}
\[\varnothing; H \vdash \text{true}.\text{if}[] := \symb{this}.>0(\symb{this}.\text{div}_{10^3}(\symb{this}.\text{income})): (A_{\beta_i}, \delta_{\beta_i}).\]
\end{small}

We can again instantiate the update rule.

\begin{small}
\begin{mathpar}
 \inferrule [Instance Type Update]
    {\varnothing; H \vdash \text{true}.\text{if}[] := \symb{this}.>0(\symb{this}.\text{div}_{10^3}(\symb{this}.\text{income})): (A_{\beta_i}, \delta_{\beta_i}) \\\\
    \symb{this}\!:\! (A_{\beta_i}, \delta_{\beta_i})\coloncolon y\!:\! (A_{\beta_i}, \delta_{\beta_i})\coloncolon \varnothing; H \vdash 1 : (A_{\beta_i}, \delta_{\beta_i})}
    {\varnothing; H \vdash \symb{this}.>0(\symb{this}.\text{div}_{10^3}(\symb{this}.\text{income})).\text{then}:= 1: (A_{\beta_i}, \delta_{\beta_i})}
\end{mathpar}
\end{small}

And a second time we instantiate {\sc Type Update} for $0$ to finally obtain

\begin{small}
\[
\begin{array}{l}
\varnothing; H \vdash \\
((\symb{this}.>0(\symb{this}.\text{div}_{10^3}(\symb{this}.\text{income}))).\text{then}
      := 1).\text{else := 0} \\: (A_{\beta_i}, \delta_{\beta_i})
\end{array}
\]
\end{small}

\subsection*{Running Example: Typing Summary}
The coercions revealed in the above steps determine the parameter $A_{\beta_i}$ in summary as follows.
\[
\begin{array}{ll}
A_{\beta_i} = & [\text{income} \mapsto H, \\
            & \ \text{div}_{10^3} \mapsto H,\\
            & \ >0 \mapsto H \\
            & \ \text{if} \mapsto H,\\
            & \ \text{ord} \mapsto S_{\text{if}}]
\end{array}
\]
I.e., the only possible instantiation for $A_{\beta_i}(\text{ord})$ is $H$. We cannot meet the required 
constraint $A_{\beta_i}(\text{ord}) = L$ necessary to call it from the outside in $\chi$ as 
explained initially. Therefore, the example configuration cannot be typed with the
constraint income $\mapsto H$.

\subsection*{Borderline Example for Confinement}
The confinement property states that remote calls can only be addressed to $L$ methods.
But does this simple security property guarantee that no hidden $H$ methods can be returned
with the reply to such a call? Consider the following example
\[
\alpha[\varnothing, [\text{leak} = \varsigma (y) \symb{this}, \text{key} = \varsigma (y) n ]]
\]
where $n$ is an integer representing a secret key. Let the security assignment for $\alpha$ be
$\{ \text{leak} \mapsto L, \text{key} \mapsto H \}$.
One may think that an activity $\beta$ could contain a call $\alpha.\text{leak}.\text{key}$ 
since the method leak is $L$ enabling the remote call to $\alpha.\text{leak}$. Once the
call result is returned into $\beta$, it would evaluate to the active object of $\alpha$ 
inside $\beta$ (since {\it this} represents this active object of $\alpha$). Since
we are now already in $\beta$, it might seem possible to apply the method key to extract the key.

How does the security type system prevent this? 
Since the typing for {\it this} inside $\alpha$ is only possible with the $PC$
as $M_\alpha = \sqcup_{i\in \{\text{leak, key}\}} A_{\alpha}(i) = H$ (since $A_\alpha(\text{key})= H$), 
the typing for {\it this} yields

\begin{small}
\[ \text{\sc Instance Val Self} \qquad \symb{this}\!:\! \Sigma_{\alpha} \coloncolon T; H \vdash {\it this}: (A_\alpha, \delta_\alpha). \]
\end{small}

Typing the object $\alpha$ must use the following instance of {\sc Type Object}.

\begin{small}
\begin{mathpar}
  \inferrule [Instance Type Object] 
    { \symb{this}\!:\! \Sigma_{\alpha} \coloncolon y\!:\! \Sigma_{\alpha}
      \coloncolon \varnothing; A_\alpha(\text{leak}) \vdash  \symb{this}: \Sigma_{\alpha} \\\\
      \symb{this}\!:\! \Sigma_{\alpha} \coloncolon y\!:\! \Sigma_{\alpha}
      \coloncolon \varnothing; A_\alpha(\text{key}) \vdash  n: \Sigma_{\alpha}}
    {\varnothing; \sqcup \{A_\alpha(\text{leak}), A_\alpha(\text{key})\}
     \vdash [\text{leak} = \varsigma (y) \symb{this}, \text{key} = \varsigma (y) n]: \Sigma_{\alpha}}
\end{mathpar}
\end{small}

Now, matching the instance of {\sc Val Self} for {\it this} with the first proviso of the
instance of {\sc Type Object} coerces $A_\alpha(\text{leak})$ to $H$ contradicting the
initial specification. I.e., the method leak is forced to be $H$ and cannot be called remotely.

\subsection*{C: Formal Semantics of \aspfun}
For a concise representation of the operational semantics, we define contexts as expressions 
with a single  hole ($\bullet$).  A context $E[t]$ denotes the term
obtained by replacing the single hole by $t$.
\[
E ::=
\begin{array}[t]{l}
  \bullet\, |\, [l_i=\varsigma(y)E,
    l_j=\varsigma(y_j)t_j^{j\in(1..n)-\{i\}}]\, |\, E.l_i(t)\, |\,  \\
   s.l_i(E)\, |\, E.l_i:=\varsigma(y)s  \,|\,s.l_i:=\varsigma(y)E |\,Active(E)
  \end{array}
    \]
This notion of context is used in the formal semantics of \aspfun in Table \ref{tab:asp_sem}
and also in the definition of visibility (see Definition \ref{def:vis}).
\begin{table*}[!ht]
\begin{mathpar}
    \inferrule [call] {l_i \in\{l_j\}^{j\in1..n}} {E\left[[l_j =
        \varsigma(y_j)s_j]^{j\in1..n}.l_i(t)\right] \loc E\left[s_i\{\symb{this}\gets[l_j
        = \varsigma(y_j)s_j]^{j\in1..n} ,y_i\gets t\}\right]}

    \inferrule [update] {l_i \in\{l_j\}^{j\in1..n}} {E\left[[l_j =
        \varsigma(y_j)s_j]^{j\in1..n}.l_i:=\varsigma(y)t\right]\loc
       E\left[[l_i=\varsigma(y)t,
        l_j=\varsigma(y_j)s_j^{j\in(1..n)-\{i\}}]\right]}

\inferrule[local]
  {s\loc s'}
  {\alpha[f_i\mapsto s \!\coloncolon\! Q,t]\coloncolon C \dist \alpha[f_i\mapsto s'\!\coloncolon\! Q,t]\coloncolon C }

\inferrule[active]
  {\gamma\notin (\dom(C)\cup\{\alpha\})\\\symb{noFV}(s) }
  {\alpha[f_i\mapsto E[\symb{Active}(s)]\!\coloncolon\! Q,t]\coloncolon C \dist \alpha[f_i\mapsto E[\gamma] \!\coloncolon\! Q,t]\coloncolon\gamma[\varnothing,s]\coloncolon C }

\inferrule[request]
  {f_k \text{ fresh}\\\symb{noFV}(s) \\ \alpha\neq\beta}
  {\alpha\left[f_i\mapsto E[\beta.l(s)]\!\coloncolon\! Q,t\right]\coloncolon\beta[R,t']\coloncolon C \dist \alpha\left[f_i\mapsto E[f_k] \!\coloncolon\! Q,t\right]\coloncolon\beta\left[f_k\mapsto t'.l(s)\!\coloncolon\! R,t'\right]\coloncolon C }

\inferrule[self-request]
  {f_k \text{ fresh}\\\symb{noFV}(s)}
  {\alpha\left[f_i\mapsto E[\alpha.l(s)]\!\coloncolon\! Q,t\right]\coloncolon C \dist \alpha\left[f_k\mapsto t.l(s)\coloncolon f_i\mapsto E[f_k]\!\coloncolon\! Q,t\right]\coloncolon C }

\inferrule[reply]
  {\beta[f_k\mapsto s\!\coloncolon\! R,t']\in\alpha[f_i\mapsto E[f_k]\!\coloncolon\! Q,t]\coloncolon C}
  {\alpha[f_i\mapsto E[f_k]\!\coloncolon\! Q,t]\coloncolon C \dist \alpha[f_i\mapsto E[s]\!\coloncolon\! Q,t]\coloncolon C }

\inferrule[update-AO]
  {\gamma \!\notin\! \dom(C)\!\cup\!\{\alpha\}\\\symb{noFV}(\varsigma(x,y)s)\\
  \beta[R,t']\!\in\!\alpha[f_i\mapsto E[\beta.l\!\!:=\!\!\varsigma(x,y)s]\!\coloncolon\! Q,t]\!\coloncolon\! C}
  {\alpha[f_i\mapsto E[\beta.l:=\varsigma(x,y)s]\coloncolon Q,t]\coloncolon C \dist \alpha[f_i\mapsto E[\gamma] \coloncolon Q,t]\coloncolon\gamma[\varnothing,t'.l:=\varsigma(x,y)s]\coloncolon C }
  \end{mathpar}  
  \caption{\aspfun semantics}
  \label{tab:asp_sem}
\end{table*}

\subsection*{D: Indistinguishability}
In \aspfun active objects are created by activation, futures by method calls. Names
of active objects and futures may differ in evaluations of the same configuration but this does not
convey any information to the attacker. In order to express the resulting structural
equivalence, we use typed bijections like \cite{bn:03} that enable the definition of an isomorphism
of configurations necessary to define indistinguishability. This technique of using the existence
of ``partial bijections'' to define an isomorphism between configurations only serves to express
equality of visible parts but is rather technical as it needs to provide differently
typed bijections for the involved structure, e.g. futures, objects, and request lists. 

\begin{definition}[Typed Bijection]
A typed bijection is a finite partial function $\sigma$ on activities $\alpha$
(or futures $f_k$ respectively) such that for a type $T$
\[ \all a: \dom(\sigma).\ \vdash a: T\ \imp\ \vdash \sigma(a): T . \]
\end{definition}
By $t[{\sigma,\tau}] =\mid_{\symb{sec}} t'$ we denote the equality of terms up to replacing all occurrences
of activity names $\alpha$ or futures $f_k$ by their counterparts $\tau(\alpha)$
or $\sigma(f_k)$, respectively, restricted to the label names in {\it sec}, i.e.,
in the object terms $t$ and $t'$ we exempt those parts of the objects that are
private.
The local reduction with $\loc^*$ of a term $t$ to a value
$t_e$ (again up to future and activity references) is written as $t \Downarrow t_e$.

\begin{definition}[Equality up to Name Isomorphism]
\label{def:ind}
An equality up to name isomorphism is a family of equivalence relations on \aspfun terms 
indexed by two typed bijections $(\sigma, \tau) := R$ and security assignment {\it sec}
consisting of the following differently typed sub-relations; 
the sub-relation's types are indicated by the naming convention: $t$ for $\varsigma$-terms, $\alpha$,
$\beta$ for active objects, $f_k, f_j$ for futures, $Q_\alpha, Q_\beta$ for request queues.
\begin{eqnarray*}
t \indR t' & \equiv & t \Downarrow t_e \wedge t' \Downarrow t'_e \wedge\ t_e[{\sigma,\tau}] =\mid_{\symb{sec}} t'_e \\[1ex]
\alpha \indR \beta & \equiv & \sigma(\alpha) = \beta \\[1ex]
f_k \indR f_j  & \equiv & \tau(f_k) = f_j \\[1ex]
Q_{\alpha} \indR Q_{\beta}\  & \equiv\ &
      \left(\begin{array}{l}
            \dom(\tau) \supseteq \dom(Q_{\alpha}) \\
            \ran(\tau) \supseteq \dom(Q_{\beta}) \\
            \all f_k \in \dom(Q_\alpha).\\
            Q_{\alpha}(f_k) \indR Q_{\beta}(\tau(f_k))
      \end{array}\right) \\ [1ex]
\end{eqnarray*}
\vspace{-7ex}
\begin{eqnarray*}
\alpha[Q_{\alpha}, t_{\beta}] \indR \beta[Q_{\beta}, t_{\alpha}] & \equiv & \alpha \indR \beta \wedge Q_{\alpha} \indR Q_{\beta} \wedge t_{\alpha} \indR t_{\beta} \\[1ex]
\end{eqnarray*}
\end{definition}
Such an equivalence relation defined by two typed bijections $\sigma$ and $\tau$ may exist
between 
given sets $V_0, V_1$ of active object names in $C, C_1$. If $V, V_1$ correspond to the viewpoints
of attacker $\alpha$ in $C$ and its counterpart in $C_1$ we call this equivalence relation 
indistinguishability.

In the following, we use the {\it visibility range} based on Definition \ref{def:vis} as
$VI_{\symb{sec}}(\alpha,C) \equiv \{ \beta \in \dom(C) \mid \beta \sqsubseteq_C^{{\symb{sec}}} \alpha \}$.
\begin{definition}[Indistinguishability]
Let $C, C_1$ be arbitrary configurations, well-typed with respect to a security specification \symb{sec}, 
active object $\alpha \in \dom(C)$ and $\alpha \in \dom(C_1)$ (we exempt $\alpha$ from renaming 
for simplicity).
Configurations $C$ and $C_1$ are called indistinguishable with respect to $\alpha$ and {\it sec}, 
if $\alpha$'s visibility ranges are the same in both up to name isomorphism.
\[ 
C \inda C_1 \equiv 
      \exists\ \sigma, \tau. \left( \begin{array}{l}
                VI_{\symb{sec}}(\alpha,C) = \dom(\sigma)\\
                VI_{\symb{sec}}(\alpha, C_1) = \ran(\sigma)\\ 
                \forall\ \beta \in VI_{\symb{sec}}(\alpha,C). \\ \ C(\beta) =_{\sigma,\tau} C_1(\sigma(\beta))
         \end{array}\right)
\]
\end{definition}

As an example for $\alpha$-indistinguishable configurations consider the running example.
In the original (non-fallacious) form, $\beta_1$.income could be $42$ in configuration $C$
and it could be $1042$ in configuration $C'$. Since income is specified as $H$ those two configurations
can be considered as $\alpha$-indistinguishable (with respect to $\beta_1$). 
Attacker $\alpha$ sees no difference between the two. In the fallacious example, however, he'd notice a difference
when calling the quicksort algorithm that implicitly drafts information from income through ord:
here, $C$ and $C'$ would be distinguishable since $\beta_1$.ord is 0 in $C$ and 1 in $C'$.

\subsection*{E: Noninterference Proof}
\it Theorem 2 (Soundness)}
For any well-typed configuration, 
we have noninterference with respect to $\alpha \in C$, \ie
\[ 
\left(
\begin{array}{c}
\vdash C: \langle \Gamma_{\symb{act}}, \Gamma_{\symb{fut}},  \symb{sec}\rangle \\
 \alpha[Q,a] \in C 
\end{array}\right)
\Rightarrow 
 \alpha{\mbox{\rm {-noninterference}}}\ C\ \symb{sec} \,.
\]
{\it Proof:}\\
Let $C_1$ be another arbitrary but fixed configuration such that $C \inda C_1$.
This means that for any $\beta \in \dom(C)$, if $\beta \in$ visibility range of $\alpha$ -- 
we have that $\sigma(\beta) \in \dom(C_1)$ and $C(\beta) =_{\sigma,\tau} C_1(\sigma(\beta))$ for some $\tau$ and $\sigma$. 
That is, aside differently named  futures (and active object references) these two activities are structurally 
the same and contain the same values.
For the sake of clarity of the proof exposition, we leave the naming isomorphism
implicit, i.e. use the same names, e.g., $\beta$, $f_k$, for both sides, i.e., 
for $\beta, \sigma(\beta)$ and $f_k, \tau(f_k)$.
Note, that
the type of the configurations $C'$ and $C_1'$ is in some cases an extension to the types of $C$ and $C_1$, 
as described in the Preservation Theorem \ref{thm:subred}.

The proof is an induction over the reduction relation combined with a case analysis whether an arbitrary 
$\beta \in \dom(C)$ is in the visibility range of $\alpha$ or not.

If, for the first case, $C \dist C'$ by some reduction according to the semantics rules in the part of the configuration
that is {\it not visible} to $\alpha$, then for most cases trivially no change becomes visible by the transition to $C'$:
for any 
local reduction, 
this is the case since the visibility relation is unchanged. Hence, ``invisible'' objects remain invisible. 
If $C \inda C_1$ and $C \dist C'$, then also $C' \inda C_1$ whereby we trivially have the conclusion since
$C_1 \diststar C_1$ (in zero steps).
This observation is less trivial for the rules {\sc request} and {\sc active} where 
new elements, futures and activities, respectively, are created. In the case of {\sc request}, 
let $\beta[f_k \mapsto E[\gamma.l(t)] :: Q_\beta, t_\beta] \in C$ and $\gamma[Q_\beta,t_\gamma] \in C$ 
with $\beta$ in the $\alpha$-invisible part and $\gamma$ visible to $\alpha$.
The fact that there are no side effects 
provides that request $f_m$ 
created in the request step in $\gamma$, i.e. $\gamma[f_m \mapsto t_\gamma.l(t):: Q_\gamma, t_\gamma]$ in $C'$, 
is not $\alpha$-visible.

Similarly, if a new activity $\gamma$ is created from a method in $\beta$ according to {\sc active}, 
then 
$\gamma$ will not be in the visibility range of $\alpha$ 
since $\beta$ was not visible to $\alpha$ by Definition \ref{def:vis} of the visibility relation. 
Thus, for $\beta$ not visible to $\alpha$, if $C \inda C_1$ and $C \dist C'$, then also $C' \inda C_1$ 
and the conclusion holds again because $C_1 \diststar C_1$. This closes the case of 
{\it non-$\alpha$-visible} reductions.

If $C \dist C'$ by some reduction in the {\it $\alpha$-visible} part, we need to consider all cases
individually as given by the induction according to the semantics rules. 

If $C \dist C'$ by semantics rule {\sc request} then  $C$ must have contained
$\beta[f_k \mapsto E[\gamma.l(t)] :: Q_\beta, t_\beta]$ and $\gamma[Q_\gamma,t_\gamma]$ for some
$\beta$, $f_k$, and $\gamma$. Hence, $\beta[f_k \mapsto E[f_m] :: Q_\beta, t_\beta]$ and
$\gamma[f_m \mapsto t_\gamma.l(t):: Q_\gamma, t_\gamma]$ in $C'$ for some new future $f_m$.
Since $\beta$ is $\alpha$-visible, so is $\gamma$ by definition of visibility
(since $f_k$ was created from $t_\beta$, $t_\beta$ must have an $L$-method containing
$\gamma$). By confinement, $f_m$ has global level $\delta_\beta$ and $l$ is $L$.
Since $C \inda C_1$, and $\beta, f_k, \gamma$ visible to $\alpha$, we 
have (up to isomorphism of names) 
that $\beta[f_k \mapsto E[\gamma.l(t)] :: Q_\beta, t_\beta]$ and $\gamma[Q_\gamma,t_\gamma]$ 
in $C_1$. Therefore, we can equally apply the rule {\sc request} to $C_1$ to obtain
that $C_1'$ contains $\beta[f_k \mapsto E[f_m] :: Q_\beta, t_\beta]$ and 
$\gamma[f_m \mapsto t_\gamma.l(t):: Q_\gamma, t_\gamma]$. In $C_1$, $\gamma$
is also $\alpha$-invisible and $l$ is typed $L$ as well.
Now, the $\alpha$-visible parts of $C_1'$ are equal to the ones of $C'$ apart from the new
future $f_m$. However, based on the future bijection $\tau$ that exists due to indistinguishability
between $C$ and $C_1$ we can extend this for $f_m$ to a bijection $\tau'$. In addition, by preservation,
$C'$ as well as $C_1'$ are well-typed whereby finally $C' \inda C_1'$ and this finishes the {\sc request}-case.
%

Another, also less obvious case for new elements in the $\alpha$-visible part, is the one for {\sc active}. 
However, here we have a very similar situation as in the {\sc request} case. If, in $C$, there is some 
$\beta[f_k \mapsto E[Active(t)] :: Q_\beta, t_\beta]$, we also have $\beta$ alike in $C_1$,
whereby we get in the next step -- according to rule {\sc active} -- $\beta[f_k \mapsto E[\gamma] :: Q_\beta, t_\beta]
\parallel \gamma[\varnothing, t]$ in $C'$ replacing 
the previous $\beta$. We can also apply {\sc active} in $C_1$ so 
that $\beta[f_k \mapsto E[\gamma] :: Q_\beta, t_\beta]
\parallel \gamma[\varnothing, t]$ in $C'$ and $C_1'$ as well instead of just the old $\beta$.
Indistinguishability is preserved since a bijection $\sigma'$ exists as extension to $\sigma$ to
the new activity $\gamma$ and by preservation again $C'$ and $C_1$ remain well-typed.
We are finished with the case for {\sc active} since $C' \inda C_1'$.

The other cases, corresponding to the remaining 
semantics rules, are of very a similar nature. 
Thus, the second part of $\alpha$-visible parts of the configurations $C$ and $C_1$ is also finished and completes
the proof of the theorem.
\hfill {$\Box$}

\end{document}